**REVIEW**

# Parallel Two-Photon Lithography: From Optical Nonlinearities to Energy Limits, Dose Control and Throughput Benchmarks

**Michalis Stavrou*, Panagiotis A. Loukakos, David Gray, and Maria Farsari**

*Institute of Electronic Structure and Laser (IESL), Foundation for Research and Technology–Hellas (FORTH), Heraklion, Crete, Greece*

**Correspondence: Michalis Stavrou (m.stavrou@iesl.forth.gr)*

**ABSTRACT**

Two-photon lithography (TPL) prints three-dimensional polymer architectures with sub-diffraction features, but writing them voxel by voxel limits its throughput. This review examines how parallelization overcomes this limit with fixed multi-focus arrays, reconfigurable holography using spatial light modulators and digital micromirror devices, hybrid metalens arrays, shaped beams, temporally focused projection and two-step absorption. An energy-deposition framework built on measured nonlinear absorption coefficients shows how the absorption order and coefficients of a resist determine its voxel size, energy per voxel and attainable voxel rate. Applied to parallel writing, the framework quantifies the costs of non-uniform foci, coherent crosstalk between foci, side-lobe losses in shaped beams and weaker axial confinement in projection. Published systems reach peak rates above $10^8$ voxels $s^{-1}$ with scanned focus arrays and up to $1.7 \times 10^9$ voxels $s^{-1}$ with grayscale projection under a fine-voxel definition, and parallel features as small as 20–55 nm. As rates rise, however, dose uniformity, proximity effects, heat, stitching and data bandwidth increasingly limit fidelity. A minimum reporting standard that includes the energy per voxel and sustained part-level rates is therefore proposed, together with priorities for the field: shared benchmark parts, predictive dose control and resists designed for parallel writing.

## 1 Introduction

### 1.1 The Throughput Limit of Serial Writing

Two-photon lithography (TPL) builds three-dimensional (3D) polymer structures with nanoscale features, but it does so one voxel at a time. Introduced by Maruo, Nakamura and Kawata in 1997, TPL uses two-photon absorption (TPA) to initiate polymerization only within a tightly focused laser beam [1]. In 2001, Kawata and co-workers printed a micrometre-sized bull sculpture with about 120 nm resolution [2], which showed the defining feature of the technique: nonlinear, thresholded photopolymerization creates features well below the diffraction limit.

TPL has since been used to print arbitrary 3D architectures from nanometres to centimetres [3, 4]. Because the local dose sets both the feature size and the degree of conversion, the technique has produced micro-optical and photonic components, architected materials, microelectromechanical systems, biomedical devices and, with grayscale exposure, four-dimensional microfabrication [4–6].

Standard TPL, however, writes point by point. Scanning a single focal volume through the resin gives precise control of the local exposure but limits the volumetric throughput, so macroscale parts can take days to print [5, 7]. For a fixed voxel geometry, packing density and scan speed, the writing time scales with the cube of the part's characteristic length [8]. A part ten times larger therefore needs a thousand times more voxels.

Overcoming this scaling requires exposing many voxels at once rather than moving a single focus ever faster. The available laser power must then be shared so that every focus still delivers the dose needed to initiate polymerization, a constraint set by the nonlinear absorption and energy deposition that govern voxel formation.

### 1.2 Physical Principles of Two-Photon Lithography

TPA was predicted by Maria Göppert-Mayer in 1931. She showed that a molecule can absorb two photons simultaneously through a virtual intermediate state, with a probability proportional to the square of the intensity, $I^2$ [9, 10] (Figure 1a). The effect was first observed in 1961, when Kaiser and Garrett detected fluorescence from $CaF_2$:$Eu^{2+}$ crystals excited by a ruby laser [10, 11]. In 1990, Denk, Strickler and Webb used it to confine fluorescence excitation to the focal volume of a laser-scanning microscope and noted its potential for localized photochemistry [12].

The efficiency of two-photon excitation is characterized by the TPA cross-section $\sigma_2$, which is expressed in Göppert-Mayer units (1 GM = $10^{-50}$ cm$^4$ s photon$^{-1}$) [13]. Because the excitation rate scales as $\sigma_2 I^2$, TPL operates at peak intensities between GW cm$^{-2}$ and TW cm$^{-2}$, where the upper end approaches higher-order absorption and photoionization [4]. These intensities are reached with ultrashort pulses, most commonly from femtosecond Ti:sapphire lasers [4, 10].

Initiators designed for TPA have cross-sections one to two orders of magnitude larger than those of conventional ultraviolet initiators, which lowers the required intensity [14, 15].

The square-law dependence is what makes 3D writing possible. In a weakly absorbing medium, the power crossing each plane along a focused beam is nearly constant, so one-photon excitation exposes the entire beam path. Under TPA, by contrast, the excitation integrated over each transverse plane falls as $1/[1 + (z/z_R)^2]$, where $z_R$ is the Rayleigh range (Figure 1b), and is therefore concentrated near the focus [10]. The squared intensity profile is also radially narrower, by a factor of $\sqrt{2}$ [16]. This confinement requires negligible one-photon absorption at the writing wavelength [4], and it weakens under wide-field illumination, which is why projection relies on temporal focusing (Section 7).

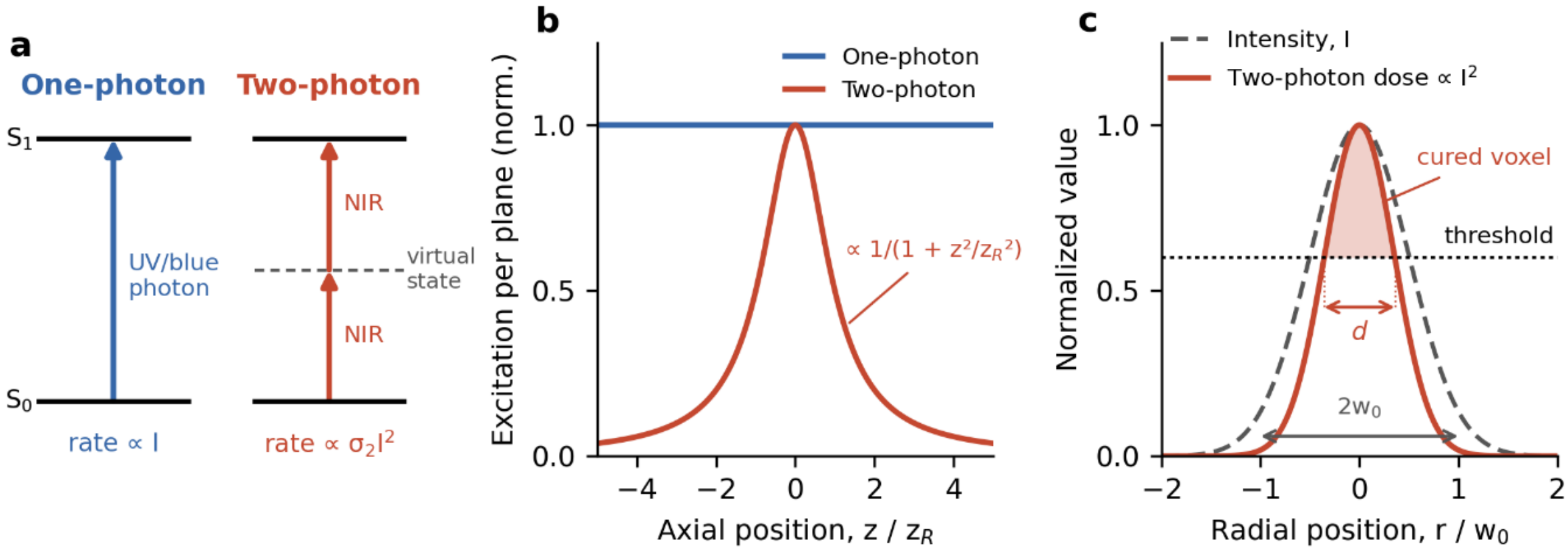

**Figure 1.** Physical principles of two-photon writing. (a) One-photon absorption excites a molecule from $S_0$ to $S_1$ at a rate proportional to the intensity *I*, whereas two near-infrared photons drive the same transition through a virtual state at a rate proportional to $\sigma_2 I^2$. (b) Excitation integrated over transverse planes along a focused Gaussian beam. It is constant for one-photon absorption but falls as $1/[1 + (z/z_R)^2]$ for two-photon absorption. (c) Radial intensity and two-photon dose at the focus. Only regions above the polymerization threshold solidify, so the voxel diameter *d* is smaller than the beam diameter $2w_0$. Original figure prepared for this review.

Photochemistry converts this localized excitation into solid material. In most resists, an excited photoinitiator undergoes intersystem crossing and generates radicals that crosslink the monomers, most often acrylates, and development then removes the unexposed resin [4, 16]. Other materials include the epoxy resist SU-8, hydrogels, composites [16] and hybrid organic–inorganic sol–gels such as the zirconium–silicon resist SZ2080™ [3], including formulations with ultralow shrinkage [17].

The material response adds a second nonlinearity, because features survive development only where the crosslink density exceeds a critical value [4, 16]. The reaction also depends on radical diffusion, viscosity, local temperature and the mobility of oxygen and other quenchers. In multifunctional acrylates, for example, polymerization is complete in less than a millisecond, and its kinetics are shaped by oxygen diffusion and inhibitor depletion [18]. Ambient temperature, humidity and stray light should therefore be controlled [4].

Thresholding allows features smaller than the focal spot (Figure 1c). Voxels shrink as the exposure approaches the threshold from above [16], and stable features well below 100 nm have been reported [4]. Typical features are nevertheless around 200 nm [19], and their size depends on the focal intensity distribution, the dose and the scan speed [5]. At the other extreme, excessive exposure causes optical damage and micro-explosions [4], which bound the process window from above.

Feature size must not be confused with resolution, which is the minimum separation at which two features remain distinguishable [4]. A line can be far narrower than the focal spot even when two such lines cannot be resolved. Conventional TPL remains subject to diffraction, and proximity effects degrade fidelity further when neighbouring spots are exposed simultaneously [7]. Inhibition schemes adapted from stimulated-emission-depletion (STED) microscopy have taken optical lithography beyond the diffraction limit [20], and their parallel implementations are discussed in Section 9.

These principles also set the throughput limit. Because every voxel must receive enough dose to exceed the effective threshold, the writing rate is bounded by the optical power delivered to the resin and by the energy required per voxel [21]. Section 2 develops this relationship quantitatively.

### 1.3 From One Focus to Many

The voxel rate can be raised either by scanning a single focus faster or by exposing many voxels at once [8]. Faster scanning does not reduce the energy required per voxel, whereas parallel writing shares the available power among many foci.

Reported rates depend strongly on how a voxel is counted (Section 3). When each projected micromirror pixel is counted as a voxel, femtosecond projection reached about $3.3 \times 10^8$ voxels $s^{-1}$ in 2019 [19]. With scanned, diffraction-limited foci, a nine-focus printer wrote $3 \times 10^{11}$ voxels at nearly $10^7$ voxels $s^{-1}$ in 2020 [22], and in 2024 a 49-focus printer reached about $10^8$ voxels $s^{-1}$ [23], then the state of the art [24]. By 2026, grayscale projection had reported up to $1.7 \times 10^9$ voxels $s^{-1}$ under a fine-voxel definition [5]. Such gains matter only if fidelity is maintained, a central theme of this review.

Each focus must still exceed the polymerization threshold, which limits the number of simultaneous foci that a given laser power can support [8]. Neighbouring foci also interact through overlapping optical fields and diffusing reactive species, which degrades fidelity [7, 25]. In addition, independent control of each focus is constrained by the speed of the modulator, its damage threshold and the minimum focal spacing [5].

Parallel approaches fall into four broad families (Figure 2). Fixed arrays split a beam into a regular grid of foci with microlens arrays or diffractive optical elements. Reconfigurable holography positions and modulates each focus with spatial light modulators (SLMs) or digital micromirror devices (DMDs) [7, 8], and hybrid systems combine fixed lens arrays with programmable illumination. Shaped beams cure an entire structural element in one exposure [26], whereas projection exposes whole patterned planes and uses temporal focusing to retain axial confinement [21]. A related group of methods replaces TPA with other nonlinear mechanisms, mainly two-step absorption [21, 27].

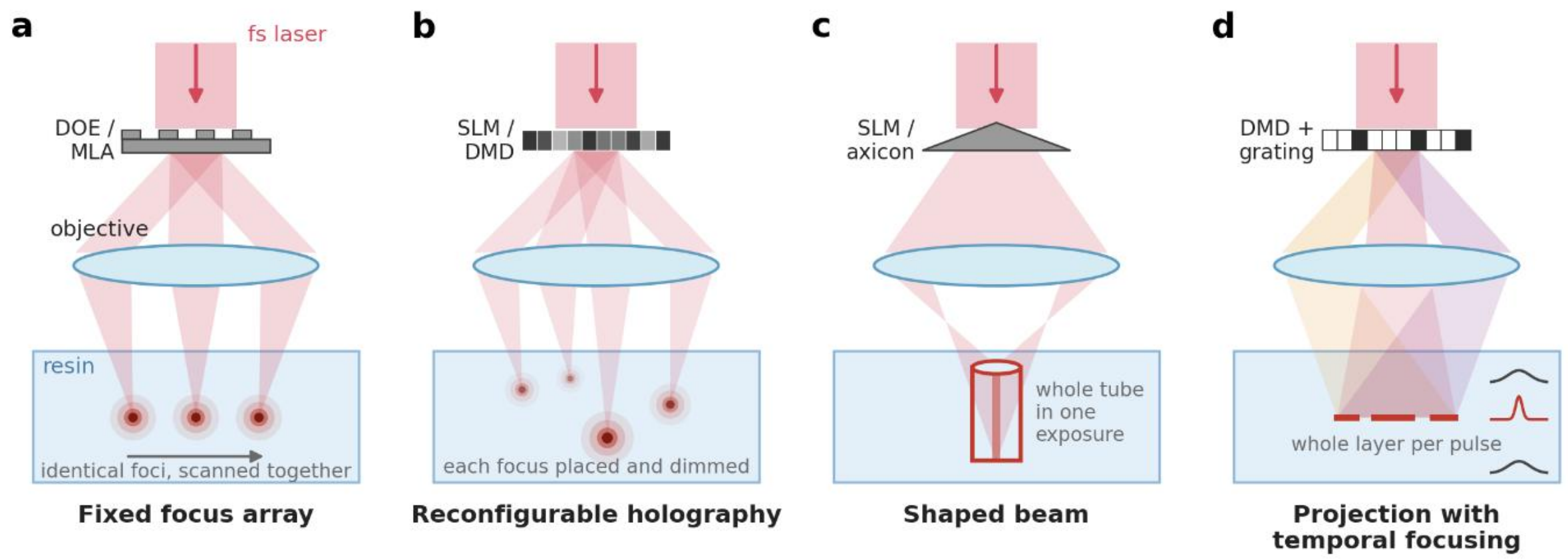

**Figure 2.** Four routes to writing many voxels at once. (a) Fixed focus arrays: a diffractive optical element (DOE) or microlens array (MLA) splits the beam into identical foci that are scanned together. (b) Reconfigurable holography: a hologram on a spatial light modulator (SLM) or digital micromirror device (DMD) places each focus in 3D and sets its power. (c) Shaped beams: a phase element forms an extended focus that cures a whole element, here a tube, in one exposure. (d) Projection with temporal focusing: a DMD pattern is imaged into the resin while the pulse spectrum is spread, and the spectral components recombine only at the focal plane, where the pulse is shortest. Original figure prepared for this review.

### 1.4 Earlier Reviews and Scope of This Review

Several reviews deal with parts of this field. They cover the physics of light-based 3D printing [21], multiphoton lithography as a method [4], new directions [28], a comparison of additive manufacturing methods by voxel rate and size [22], photoresist sensitivity as a limit on speed [29], lithography beyond the diffraction limit [20] and parallel laser micro- and nanoprocessing for functional devices [30].

Closest to the present work, Balena et al. reviewed high-speed and holographic two-photon writing, covering both parallel fabrication from a single beam and engineered intensity distributions for single-exposure patterning [26]. Other reviews treat dynamic beam shaping in ultrafast laser processing [31–33] and adaptive optics for aberration correction and parallel writing [34].

The present review covers parallelization across all families, namely fixed arrays, reconfigurable and hybrid holography, shaped beams, projection and two-step absorption. It includes developments since 2023, such as metalens arrays with more than 120,000 foci, grayscale and high-repetition-rate projection and learned hologram design. It also develops a common energy-deposition framework for all families (Section 2) and uses it to argue for standardized throughput benchmarks (Section 3).

Section 2 develops the energy-deposition model, the costs of parallel writing and a worked example based on measured coefficients, and Section 3 examines throughput metrics. Sections 4–8 review these approaches in turn, from fixed arrays to routes beyond TPA, and Sections 9–12 cover hardware and materials, cross-cutting challenges, applications and a comparison of approaches. Section 13 identifies open questions, and Section 14 summarizes the conclusions.

## 2 Voxels, Dose, and the Throughput Ceiling

The speed limit of TPL is set by the physics of voxel formation, not only by how fast the optics can move a focus. This section quantifies the two principles of Section 1.2, quadratic excitation and a dose threshold, and derives the constraints that every parallelization strategy must satisfy. The model is deliberately simple, with Gaussian foci, a sharp threshold and a local, instantaneous response, so that the scaling laws remain transparent. Its limitations are stated at the end of Section 2.1.

### 2.1 Nonlinear Energy Deposition

Following Stavrou et al. [35], the absorption coefficient of the resist can be expanded as a power series in the local intensity $I$:

$$\alpha(I) = \alpha_1 + \alpha_2 I + \alpha_3 I^2 + \alpha_4 I^3 + \cdots \tag{1}$$

Here $\alpha_1$ is the linear absorption coefficient, and $\alpha_2$, $\alpha_3$ and $\alpha_4$ (often written as $\beta$, $\gamma$ and $\delta$) are the two-, three- and four-photon absorption coefficients. Because $dI/dz = -\alpha(I)I$, the power absorbed per unit volume is

$$q = \alpha(I)\,I = \sum_n \alpha_n\,I^n \tag{2}$$

At the writing wavelength, the resist is nearly transparent ($\alpha_1 \approx 0$), and the lowest order $n$ whose combined photon energy reaches an absorbing state dominates. For a paraxial Gaussian focus with waist $w_0$, Rayleigh range $z_R$ and peak intensity $I_0$, carried by Gaussian pulses of duration $\tau$ (full width at half maximum), integrating Equation (2) over one pulse gives the energy deposited per unit volume by an $n$-photon process:

$$\varepsilon_n(r,z) = \alpha_n \tau_n I_0^n \left(1 + \frac{z^2}{z_R^2}\right)^{-n} \exp\left[-\frac{2nr^2}{w_0^2(1 + z^2/z_R^2)}\right], \qquad \tau_n = \tau\sqrt{\frac{\pi}{4n\ln 2}} \tag{3}$$

The coefficient $\alpha_n$ characterizes the bulk formulation. It is related to the molecular cross-section $\sigma_n$ through the photon energy $h\nu$, Avogadro's number $N_A$ and the molar concentration $\rho$ of the absorbing molecules [35]:

$$\sigma_n = \frac{\alpha_n (h\nu)^{n-1}}{N_A\,\rho} \tag{4}$$

For $n = 2$, this expression gives the TPA cross-section $\sigma_2$ of Section 1.2. In a mixture, each species contributes a term $N_A\rho\sigma_n/(h\nu)^{n-1}$ to $\alpha_n$, so the deposited energy is shared between photoinitiator and matrix according to their concentrations and cross-sections.

In our previous work, we measured these coefficients by open-aperture Z-scan with 80 fs pulses for the hybrid resist SZ2080™, both neat and with 1 wt% of Irgacure 369 (IRG369) or 4,4′-bis(diethylamino)benzophenone (BIS) (Table 1) [35]. Neat SZ2080™, whose lowest allowed transitions lie near 260 nm, absorbs through two, three and four photons at 515, 800 and 1030 nm, respectively. Adding a photoinitiator raised $\beta$ at 515 nm about sixfold (IRG369) or thirteenfold (BIS) and lowered the dominant order at longer wavelengths. IRG369 shifted absorption at 1030 nm from four to three photons, whereas BIS made two-photon absorption dominant at all three wavelengths, an effect that we tentatively attributed to a donor–acceptor charge-transfer complex with the resin [35].

**Table 1.** Dominant absorption order and effective nonlinear absorption coefficients of SZ2080™ formulations, measured by open-aperture Z-scan with 80 fs pulses [35]. 2PA, 3PA and 4PA denote two-, three- and four-photon absorption. $\beta$ is given in $10^{-13}$ cm W$^{-1}$, $\gamma$ in $10^{-25}$ cm$^3$ W$^{-2}$ and $\delta$ in $10^{-37}$ cm$^5$ W$^{-3}$, so coefficients of different order cannot be compared directly. The Rayleigh ranges were 9.8, 1.6

and 2.7 mm at 515, 800 and 1030 nm, and the values in parentheses were obtained with an approximately tenfold shorter Rayleigh range.

| **Formulation** | **515 nm** | **800 nm** | **1030 nm** |
|---|---|---|---|
| Neat SZ2080™ | 2PA, $\beta$ = 4.5 (3.0) | 3PA, $\gamma$ = 67.2 (15.7) | 4PA, $\delta$ = 108 (24.3) |
| SZ2080™ + 1 wt% IRG369 | 2PA, $\beta$ = 26.5 (11.5) | 3PA, $\gamma$ = 86.6 (21.3) | 3PA, $\gamma$ = 12.6 (2.6) |
| SZ2080™ + 1 wt% BIS | 2PA, $\beta$ = 58 (19.8) | 2PA, $\beta$ = 36.2 (13.7) | 2PA, $\beta$ = 13.5 (3.1) |

Even at 1 TW cm$^{-2}$, these coefficients attenuate the on-axis peak by at most about $10^{-3}$ over a 1 µm focal length, so beam depletion is negligible. The prepolymer, which makes up about 99 wt% of the formulation, also absorbs, and neat SZ2080™ polymerizes without any photoinitiator [35].

Integrating Equation (3) over each transverse plane shows how the absorption order sets the axial confinement:

$$\int_0^\infty \varepsilon_n \; 2\pi r \, \mathrm{d}r = \frac{\pi w_0^2}{2n} \, \alpha_n \tau_n I_0^n \left(1 + \frac{z^2}{z_\mathrm{R}^2}\right)^{1-n} \qquad (5)$$

For $n$ = 1, every plane absorbs the same energy. For $n$ = 2, the deposition falls as $1/(1 + z^2/z_\mathrm{R}^2)$ (Figure 1b), and each additional photon steepens the decrease. At fixed fluence, $\varepsilon_n$ also scales as $\tau^{1-n}$, so pulses stretched away from the focal plane deposit less energy, which is the basis of temporal focusing (Section 7). Integrating along the beam axis gives the on-axis deposited energy per unit area:

$$\int_{-\infty}^{\infty} \varepsilon_n \,(0, z) \, \mathrm{d}z = C_n \, \alpha_n \tau_n I_0^n z_\mathrm{R}, \qquad C_n = \frac{\sqrt{\pi} \; \Gamma\left(n - \frac{1}{2}\right)}{\Gamma(n)} \qquad (6)$$

with $C_2 = \pi/2$, $C_3 = 3\pi/8$ and $C_4 = 5\pi/16$. In a resist thicker than the focal region, the on-axis deposited energy therefore grows in proportion to $z_\mathrm{R}$ at fixed peak intensity. This explains the focusing dependence of our measurements [35]. A tenfold longer $z_\mathrm{R}$ raised the effective coefficients 1.5- to 5-fold, and lines written at 0.45 NA required 0.4–0.7 times the threshold intensity of lines written at 1.4 NA. The measured coefficients are effective values for a given focusing geometry and approach intrinsic material values under loose focusing [35].

Equations (1)–(6) assume an undistorted focus and a local, instantaneous response. The Kerr effect is too weak to reshape high-NA foci [35], but free carriers add absorption at high intensity, heat accumulates at printing repetition rates, and the effective nonlinearity in lithography can depend on the repetition rate [36]. Coefficients measured by Z-scan at 10 Hz therefore predict printing trends rather than exact thresholds [35].

### 2.2 Voxel Size and Uniformity

A threshold model links the deposited energy to voxel formation. If the resist polymerizes wherever the accumulated dose exceeds a threshold $D_{th}$, and $M$ pulses produce a peak dose $D_0 = M\alpha_n\tau_n I_0^n$, Equation (3) gives the voxel diameter and length:

$$d = w_0\left[\frac{2}{n}\ln\frac{D_0}{D_{\mathrm{th}}}\right]^{1/2}, \qquad l = 2z_{\mathrm{R}}\left[\left(\frac{D_0}{D_{\mathrm{th}}}\right)^{1/n} - 1\right]^{1/2} \tag{7}$$

For $n = 2$, these expressions reduce to $d = w_0[\ln(D_0/D_{th})]^{1/2}$ and $l = 2z_R[(D_0/D_{th})^{1/2} - 1]^{1/2}$. At fixed $D_0/D_{th}$, the diameter scales as $n^{-1/2}$, as does the length near threshold, so the aspect ratio $l/d$ approaches $\sqrt{2}\, z_R/w_0$ independently of the absorption order. Higher orders do not guarantee finer features, because they are usually exploited at longer wavelengths. Lines written at 1030 nm were 1.3–1.5 times wider than at 515 nm, whereas at a given wavelength and numerical aperture the width changed little with the photoinitiator or the absorption order [35]. Near threshold, mechanical stability and development, which the model omits, also set the narrowest stable line.

The model reproduces the measured trend that voxel diameter and length both vanish as the peak dose approaches the threshold [37], which is why the finest features are written just above it. Real voxels depart from this behaviour through the diffusion of radicals, oxygen and quenchers [4, 18] and through dose added by neighbouring exposures [7]. The model nonetheless shows why parallel writing is so sensitive to uniformity. Because the dose scales as $I^n$, a focus that is 3% too bright delivers about 6% more dose under TPA. At twice the threshold dose, this changes the voxel diameter by only about 4%, but at 15% above threshold, which is typical of fine-feature writing, it changes the diameter by about 19%, consistent with the 20% feature-size variations observed for 3% intensity differences between holographic foci [38] (Section 10). More generally, the dose error is about $n$ times the intensity error, so at the same 15% dose margin a 3% error changes the diameter by +28% or −41% for $n = 3$ and by +36% or −64% for $n = 4$. This order dependence holds only at equal dose margins. In terms of intensity, the width $d = w_0[2\ln(I_0/I_{th})]^{1/2}$ does not depend on $n$ [36], so at equal intensity margins, and hence equal feature sizes, all orders are equally sensitive.

Voxels are also elongated, and increasingly so at lower numerical aperture. They are about 7 µm long but less than 1 µm wide at 0.75 NA [8], compared with about 500 nm by 200 nm at 1.4 NA [23]. Elongation can be useful, because a single traced line can form a tall wall, but it limits the axial resolution.

The dose also depends on the surroundings and the exposure history. Arnoux et al. attribute such spatial and temporal proximity effects mainly to overlapping point-spread functions and radical diffusion [7], and measurements with SLMs likewise indicate a strong contribution from molecular diffusion [25]. These effects influence every writing strategy, but they matter most when many nearby foci are exposed at once.

### 2.3 Power Budget and the Serial Ceiling

These constraints define a power budget, in which throughput can increase only if more power reaches the resist or the available power is used more efficiently [21]. In the 7 × 7 focus printer of Kiefer et al. (Section 4), each focus received about 19.5 mW at 80 MHz [23]. At 1 m $s^{-1}$ and a voxel size of about 200 nm, each voxel is exposed for about 200 ns, which corresponds to about 16 pulses of 0.24 nJ, or about 4 nJ per voxel.

If the dose accumulates additively over $M$ pulses, a voxel forms at the threshold intensity $I_{th} = [D_{th}/(M\alpha_n\tau_n)]^{1/n}$ (Equations (3) and (7)), and each focus requires a pulse energy of about $(\pi w_0^2/2)\tau_1 I_{th}$, with $\tau_1 \approx 1.06\tau$. A laser pulse energy $E_p$ delivered with efficiency $\eta$ can therefore drive at most about $\eta E_p/[(\pi w_0^2/2)\tau_1 I_{th}]$ foci. Dividing a pulse among $N$ foci at fixed total power scales the total nonlinear deposition as $N^{1-n}$, and foci that fall below threshold stop polymerizing. A larger $\alpha_n$ lowers $I_{th}$ and allows more foci per laser. Because $D_{th}$ also depends on the efficiency of radical generation, however, thresholds must still be measured. Both formulations containing a photoinitiator had lower thresholds than neat SZ2080™, and BIS gave the lowest [35].

The same model explains the focusing results of Section 2.1. At a fixed scan speed, a 0.45 NA focus about three times wider than a 1.4 NA focus collects roughly three times as many pulses. Because $I_{th} \propto M^{-1/n}$, this alone lowers the threshold intensity to 0.58, 0.69 and 0.76 of its value for $n$ = 2, 3 and 4, which partly explains the measured ratios of 0.4–0.7 at 5 mm $s^{-1}$ [35]. By our estimate, the wider focus, with about ten times the area, needs four to seven times more pulse energy, but it cures a much larger volume, and the line widths increased about 4.5-fold [35]. Low numerical aperture thus trades resolution for processing capacity, which suits large structures such as scaffolds and microrobots, whereas the finest features still need high numerical aperture.

Parallel systems also differ in how they deliver energy over time. With a megahertz oscillator, each voxel accumulates dose from many weak pulses, whereas with a kilohertz amplifier a single, much stronger pulse may form it. Ouyang et al. partly attributed their 90 nm resolution to this single-pulse regime [15]. The two regimes differ in heat accumulation, radical diffusion between pulses and the timing of neighbouring exposures, and they show different effective nonlinearities [36], so performance does not transfer directly between them. Print speed also cannot be raised indefinitely by laser power alone, and more efficient use of light and more sensitive resists must contribute [24].

Within these constraints, serial TPL has two main routes to higher throughput. The first is faster scanning. Galvanometer mirrors move the focus much faster than translation stages, but only within the field of view of the objective, so large parts require stitching [8]. Synchronizing the stages with the galvanometers avoids stitching for mesoscale objects [39], and an 8 kHz resonant scanner reached 66 m $s^{-1}$, producing 12,078 tissue-engineering microscaffolds per hour, 57 times the throughput of a two-galvanometer system [40].

The second route is a more sensitive resist, because faster scanning shortens each exposure. Comparing photoresists by their threshold power at a given focus velocity, Kiefer et al. found that the advantage of the photoinitiator DETC grows with scan speed [29]. Because $I_{th} \propto M^{-1/n}$ and $M \propto f/v$, the threshold power of an ideal two-photon resist rises as $\sqrt{v}$, but that of DETC, an effective three-photon initiator [29, 36], rises only as $v^{1/3}$. Thresholds measured at two scan speeds give the effective order, $n = \Delta\ln v/\Delta\ln P_{th}$.

Nevertheless, a single focus can write only a finite number of voxels per second. At 1 m $s^{-1}$ and a 100 nm pitch, it addresses about $10^7$ voxel positions per second. Hahn et al. therefore characterized 3D printing methods by their maximum voxel rate and minimum voxel size [22]. If the laser can sustain $N$ foci, parallel writing raises the voxel rate about $N$-fold without faster scanning, at a cost in optical efficiency, uniformity, crosstalk and independent control that Section 2.4 quantifies.

## 2.4 Energy Deposition in Parallel Writing

Applied to each focus of a parallel printer, Equations (3)–(7) quantify the main constraints on throughput and show how they differ between strategies.

The first constraint is energy. Writing a voxel of diameter $d$ requires an incident energy $E_{vox}$, and for $N$ foci, repetition rate $f$, average power $P_{avg}$ and delivery efficiency $\eta$, the voxel rate $R_{vox}$ is bounded by

$$E_{\mathrm{vox}} = \frac{\pi w_0^2}{2}\,\tau_1\,M^{1-1/n}\left(\frac{D_{\mathrm{th}}}{\alpha_n \tau_n}\right)^{1/n} e^{d^2/2w_0^2}, \qquad R_{\mathrm{vox}} \le \min\left(\frac{\eta P_{\mathrm{avg}}}{E_{\mathrm{vox}}}, \frac{Nf}{M}\right) \tag{8}$$

The first bound is set by energy and the second by the number of foci that each pulse can feed. For the 7 × 7 printer, both bounds give about 2.5 × $10^8$ voxels $s^{-1}$, close to the reported 1.4 × $10^8$ voxels $s^{-1}$ [23]. For the 2000-focus holographic printer, the second bound alone gives the reported 2 × $10^6$ voxels $s^{-1}$ [15].

Equation (8) rewards three choices. First, because $E_{vox}$ grows as $M^{1-1/n}$, one strong pulse per voxel needs less energy than many weak ones. Under TPA, the 16 pulses per voxel of the 7 × 7 printer need four times the single-pulse energy at threshold. Lithography experiments support this accumulation. In acrylate resists with IRG369 or Irgacure 819, thresholds between 100 kHz and 80 MHz follow two-photon accumulation over pulses, whereas below about 10 kHz photoionization raises the effective order, to 3.5–4.5 at 4 kHz [36]. The order $n$ should therefore be read as the effective order at the operating repetition rate, which, if anything, favours single pulses further. Single-pulse writing, as in holographic printing [15] and grayscale projection [5], is thus energy-efficient. At kilohertz rates, however, it shifts the limit to the second bound, which demands very many foci per pulse (Section 9), and it raises the peak intensity by $M^{1/n}$, closer to damage. Second, $E_{vox} \propto \tau^{1-1/n}$ favours short pulses, such as the 35 fs pulses of Kim and Saha [5]. Third, $E_{vox} \propto (D_{th}/\alpha_n)^{1/n}$, so a tenfold more sensitive resist raises the energy-limited rate only about 3.2-fold for $n$ = 2. High-cross-section initiators [15] and formulations such as those in Table 1 [35] act through this term.

The second cost is crosstalk. Foci split from one pulse are mutually coherent, so their fields add before the nonlinearity acts. Midway between two equal foci separated by $s$ and with relative phase $\varphi$, the dose relative to the peak dose $D_0$ of one focus is

$$\frac{D_{\text{gap}}}{D_0} = [2(1+\cos\varphi)]^n \exp\left(-\frac{ns^2}{2w_0^2}\right) \tag{9}$$

whereas two sequential exposures leave only $2\exp[-ns^2/(2w_0^2)]$. Foci in quadrature ($\varphi = \pi/2$) add like intensities and give $2^{n-1}$ times the sequential dose. In phase, the gap dose rises by a further factor of $2^n$, whereas in antiphase it cancels on the midplane at the cost of slightly weaker, displaced foci. Near threshold ($D_0/D_{th}$ = 1.15), the gap between foci in quadrature stays uncured only for $s \gtrsim 1.2w_0$, regardless of the absorption order. Sequential writing, by contrast, allows $0.91w_0$ for $n$ = 2 and $0.65w_0$ for $n$ = 4, and in-phase foci need about $1.8w_0$ because of their overlap out of focus. Simultaneous exposure thus erases most of the packing advantage of higher-order absorption and turns the phase of each focus into a design variable. Delaying neighbouring foci by more than the pulse duration removes the cross terms altogether and restores the sequential limit, as time multiplexing does in multifocal multiphoton microscopy [41].

Large arrays impose a further limit. Out of focus, the light of all foci merges into a field whose mean intensity, $\pi w_0^2 I_0/(2p^2)$ for a square array of pitch $p$, must stay below $I_{th}$. Near threshold, this condition requires $p \gtrsim 1.3w_0$. These terms correspond to the overlap of point-spread functions that Arnoux et al. identified alongside radical diffusion [7], and they help explain why holographic foci cannot be packed arbitrarily closely [5]. In projection, where neighbouring spots overlap by design, they make the dose of each spot depend on how many of its neighbours are lit, so dense regions over-cure and sparse ones under-cure (Section 7).

The third cost is uniformity. Linearizing Equation (7) shows that keeping the voxel diameter within a fraction $\delta$ requires the focus intensities to agree within $(2/n)\ln(D_0/D_{th})\delta = 2\ln(I_0/I_{th})\delta$. For $\delta = 5\%$ at a dose 15% above threshold, this tolerance is 0.7% for $n = 2$ and 0.35% for $n = 4$, relaxing to 3.5% for two-photon writing at twice the threshold dose. At a given feature size, however, it is the same for every order (Section 2.2). Real arrays fall short. At twice the threshold, the ±13% power spread across the 49 foci of Kiefer et al. [23] would make the diameters vary from 23% below to 16% above nominal. Printers therefore either operate further above threshold, giving up the finest features, or measure and correct each focus [38] (Section 10).

Shaped beams pay a different price. The first side lobe of an ideal zeroth-order Bessel beam reaches 16% of the core intensity and therefore receives only 2.6% of the core dose under TPA and 0.07% under four-photon absorption, so isolated rings rarely cure. Each ring, however, carries about 25% more power than the core. A beam with 10–30 rings therefore holds only 2.5–7% of its power in the core, and under TPA its rings absorb 47–56% of the deposited energy, a share that grows with the number of rings, compared with less than 10% under three-photon absorption. This energy is wasted, and when a shaped beam is scanned or repeated, the ring doses of neighbouring exposures add up and complicate dose control (Section 6).

Projection with temporal focusing confines deposition less tightly than a scanned point focus. For uniform illumination at fixed fluence, Equation (3) makes the deposited energy scale as $\tau^{1-n}$. If the pulse lengthens away from the temporal focal plane as $\tau(z) \approx \tau_0(1 + z^2/z_T^2)^{1/2}$, where $z_T$ is the axial range over which it stays short, the energy deposited per pulse at a distance $z$ is

$$\varepsilon_n^{\mathrm{TF}}(z) \propto \left(1 + \frac{z^2}{z_{\mathrm{T}}^2}\right)^{(1-n)/2} \tag{10}$$

which has half the exponent of Equation (5). Under TPA, the dose away from the focal plane therefore falls only as $1/z$, compared with $1/z^2$ for a layer filled by a scanned focus, so each projected layer leaves a background dose in its neighbours that adds to the proximity effects discussed in Section 7. Higher orders steepen this decay, broader bandwidths shorten $z_T$ (Section 9) and so reduce its amplitude, and line illumination restores a steeper decay even under TPA. Grayscale control inherits the nonlinearity too. Under TPA, the 10:1 intensity range of the 32 motifs of Kim and Saha spans a 100:1 range of dose [5].

Two-step absorption fits the same framework on a different time scale, and it is the only route discussed here for which $\alpha_1$ is not negligible. Its first step is weak linear absorption that does not cure, so the threshold applies only to the second step. In a simple rate-equation picture, for exposures longer than the intermediate-state lifetime $\tau_{int}$ and for $\sigma_e \gg \sigma_g$, the second step yields a quadratic term with an effective coefficient $\alpha_{2,eff} \approx N_A \rho \sigma_g \sigma_e \tau_{int}/(h\nu)$, where $\sigma_g$ and $\sigma_e$ are the ground- and intermediate-state cross-sections. Because a real intermediate state lives far longer than the virtual state of TPA, inexpensive continuous-wave lasers suffice, with about 100 µW per focus [27] (Section 8). For dwell times shorter than $\tau_{int}$, however, the intermediate population cannot build up, and the required power grows in proportion to the scan speed rather than its square root. The lifetime therefore caps the single-focus speed and favours parallel foci. Near this limit, the effective nonlinearity also falls at faster scan speeds, which degrades print quality [42], and the same memory couples foci exposed within $\tau_{int}$ of one another. Figure 3 maps these scalings for each family.

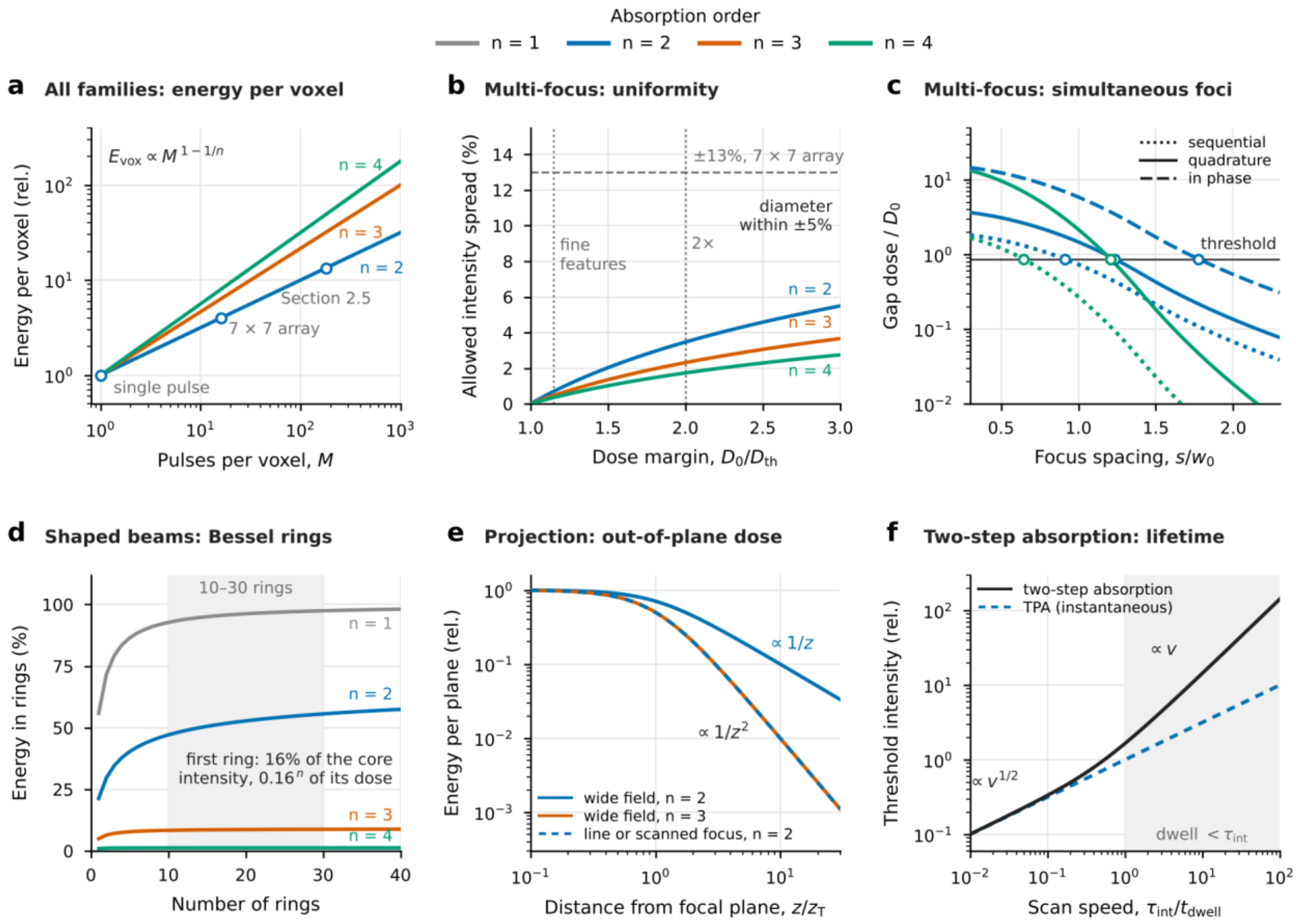

**Figure 3.** Design map of the energy-deposition model, with one panel for each issue in Section 2.4. Colours denote the absorption order $n$. (a) Energy per voxel relative to single-pulse writing, $E_{vox} \propto M^{1-1/n}$ (Equation (8)), with the cases discussed in the text marked. (b) Intensity agreement between foci that keeps the voxel diameter within ±5%, compared with the ±13% power spread of a 7 × 7 array [23]. (c) Peak dose midway between two foci, including out-of-focus overlap, for sequential exposure and for simultaneous foci in quadrature or in phase (Equation (9)). Circles mark the smallest spacing that leaves the gap uncured at $D_0 = 1.15D_{th}$: $0.91w_0$ ($n = 2$) and $0.65w_0$ ($n = 4$) for sequential exposure, $\approx 1.2w_0$ in quadrature and $\approx 1.8w_0$ in phase. Large arrays also need a pitch of $\approx 1.3w_0$ or more. (d) Share of the deposited energy absorbed in the rings of a zeroth-order Bessel beam. (e) Energy deposited in a plane at distance $z$ from the focal plane for wide-field temporal focusing (Equation (10)) and for line illumination or a scanned focus, taking $z_R = z_T$. (f) Threshold intensity versus scan speed for two-step absorption, from rate equations, and for TPA, normalized at low speed. Original figure prepared for this review.

## 2.5 A Worked Example: Why the Coefficients Matter

The coefficients make the model quantitative. For neat SZ2080™ at 515 nm, polymerization starts at about 0.25 TW cm$^{-2}$ with 240 fs pulses at 0.1 MHz, 100 µm s$^{-1}$ and 1.4 NA, and breakdown sets in near 1.8 TW cm$^{-2}$ [43]. A scanned focus with $w_0 \approx 0.2$ µm delivers $M = fw_0\sqrt{\pi}/(2v) \approx 180$ effective pulses, so Equation (3) with $\beta = 3.0$–$4.5 \times 10^{-13}$ cm W$^{-1}$ (Table 1) gives a threshold dose of $D_{th} \approx 0.6$–$0.9$ J cm$^{-3}$. Unlike a threshold power, this dose can be transferred to other writing conditions. At 5 mm s$^{-1}$ and 0.2 MHz, 25 times fewer pulses raise the threshold fivefold, to about 1.2 TW cm$^{-2}$, just below the 1.5–2 TW cm$^{-2}$ used for high-fidelity printing under those conditions [35]. With one pulse per voxel, as in kilohertz holography and projection, neat SZ2080™ would need about 3.3 TW cm$^{-2}$ for $n$ = 2, well beyond breakdown. An effective order of three, plausible at kilohertz rates (Section 2.4), would lower this to about 1.4 TW cm$^{-2}$, below breakdown, so single-pulse thresholds must be measured rather than extrapolated. BIS, whose $\beta$ is 7–13 times larger, reaches the same dose at no more than 0.9–1.3 TW cm$^{-2}$, or about 0.2 nJ per focus, which lies inside the processing window. Single pulses at kilohertz rates would also avoid the cumulative heating that degraded BIS prints at high repetition rates [35, 43].

Because the photoinitiator sets the absorption order (Table 1), the coefficients also decide which parallel architecture a resist suits. For many simultaneous foci, the lowest order is preferable. Unless the foci are time-multiplexed, interference removes the packing advantage of higher orders (Section 2.4), and a ±13% power spread between foci becomes a dose spread of +28%/−24% under TPA but +63%/−43% under four-photon absorption. For wide-field projection with temporal focusing, a higher order is preferable, because the background dose then falls as $1/z^2$ under three-photon absorption rather than as $1/z$ (Equation (10)), albeit at the cost of higher intensity and smaller fields. At 1030 nm, BIS therefore suits multi-focus writing with high-power fibre lasers, whereas IRG369 and the neat resist retain three- and four-photon confinement for projection. Knowledge of the absorption order and coefficients, combined with a threshold measured under the focusing and pulse conditions of the target printer, is thus essential for modelling energy deposition in parallel writing.

## 3 How Throughput Is Measured

Parallel TPL studies report throughput in different units and under different definitions, which makes headline values hard to compare.

The most common metric is the voxel rate, the number of voxels exposed per second [5, 21], which suits scanning systems but depends on voxel definitions that vary between studies. The other widely used metric is the volume rate, typically given in mm$^3$ h$^{-1}$. For projection, Saha et al. distinguished volumetric throughput, set by the projected area, layer thickness and time between layers, from areal throughput, set by the projected area and exposure time [19]. Reported volume rates range from about 3.9 mm$^3$ h$^{-1}$ for a millimetre-scale lattice [44] to about 215 mm$^3$ h$^{-1}$ for grayscale projection with a 113 µm field, 5000 layers per second and 1.2 µm layer spacing [5]. An interspersing projection method exceeded 0.5 mm$^2$ s$^{-1}$ per layer [45].

The voxel definition alone can shift a reported rate by more than an order of magnitude. Kim and Saha reported $7.9 \times 10^7$ voxels $s^{-1}$ when each projected focal spot was counted as one voxel. Defined by the finest printed feature, however, about 107 voxels fit within each spot, and because the finest nanowire took five pulses, the same experiment yields $1.7 \times 10^9$ voxels $s^{-1}$ [5]. Saha et al. reported about $3.3 \times 10^8$ voxels $s^{-1}$ by counting each projected micromirror pixel (about 151 nm in the resist) exposed during a 3 ms projection, whereas the corresponding 3D printing rate was 5–20 $mm^3$ $h^{-1}$ [19]. Even within one scanning experiment, Kiefer et al. reported $1.4 \times 10^8$ voxels $s^{-1}$ for diffraction-limited voxels defined by a two-photon Sparrow criterion, but $7.1 \times 10^7$ voxels $s^{-1}$ for the voxel size averaged over their 49 foci [23]. Each value is valid under its own definition, but a ranking requires the voxel dimensions and counting method, which many studies do not report.

Part geometry matters just as much. The volume rate for a solid block is not comparable with that for a sparse lattice, in which most of the swept volume remains unpolymerized. Ouyang et al. noted that random-access holographic writing is most efficient for fill fractions of 1–12%, and they compared resolution against both voxel and volume rates (Figure 4a,b) [15]. Similarly, woodpiles and line arrays are faster to print than dense, aperiodic structures with overhangs.

For manufacturing, part-level throughput, such as 12,078 microscaffolds per hour [40] or more than 50 million microparticles per day [46], is often more informative, because it includes overheads such as stage motion and stitching.

Throughput is meaningful only together with quality, because delivering a controlled dose to every voxel becomes harder as rates increase [5]. Kim and Saha therefore paired voxel and volume rates with the finest printable feature and the smallest printable pore (Figure 4c) [5]. Similarly, Hahn et al. plotted the maximum voxel rate against the minimum voxel size (Figure 4d) [22], a comparison that Somers et al. keep updated online [21].

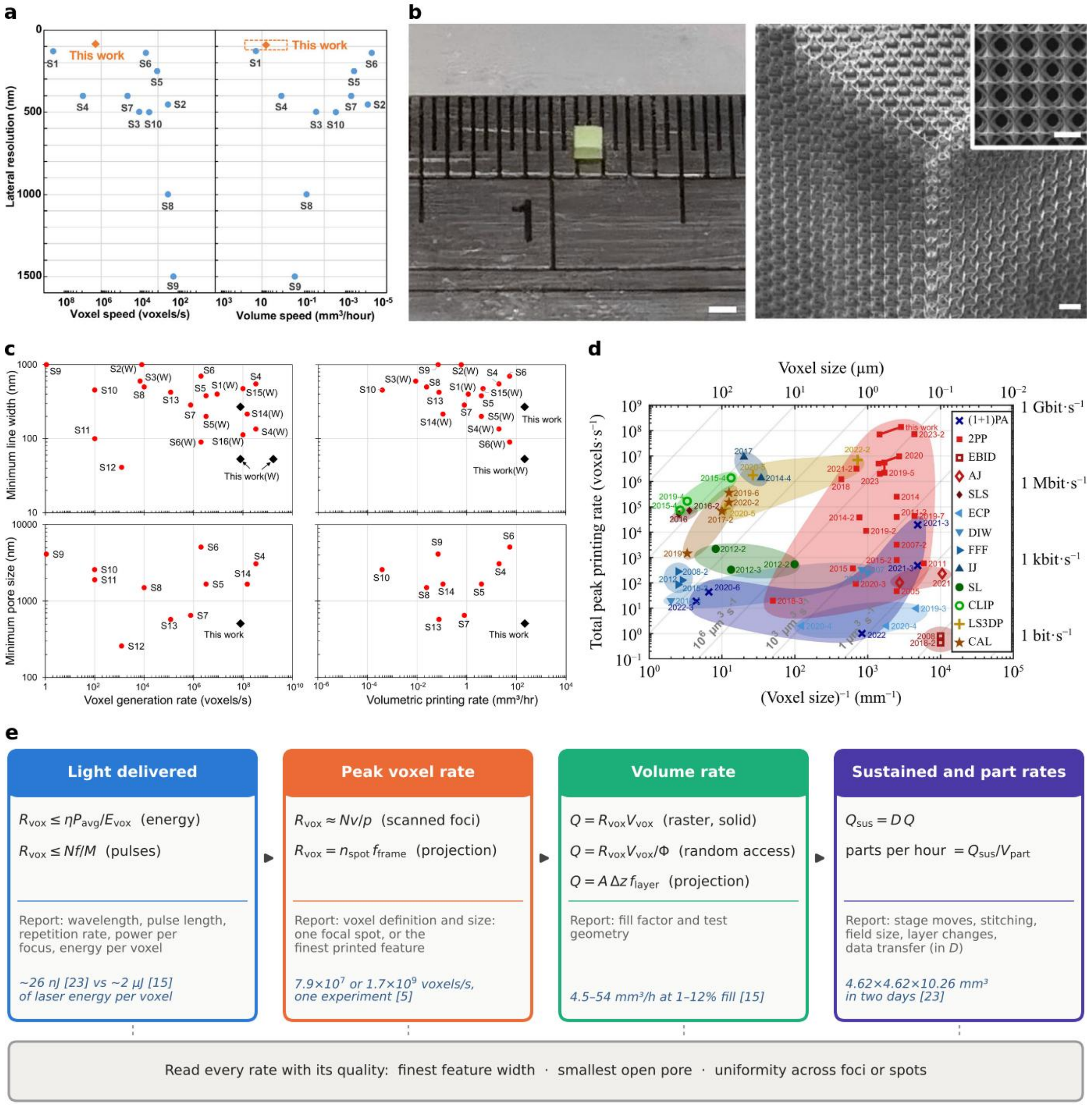

**Figure 4.** How throughput is reported. (a) Lateral resolution versus voxel rate (left) and volume rate (right) for holographic printing with up to 2000 foci ("This work") and earlier TPL studies (S1–S10 in the original source). (b) A 1.17 mm$^3$ cube of octahedral trusses printed with 100 foci, shown on a ruler, and an electron micrograph with a top-view inset. (c) Minimum line width (top) and pore size (bottom) versus voxel rate (left) and volume rate (right) for grayscale projection TPL ("This work") and earlier studies (S1–S16 in the original source). (d) Peak printing rate versus voxel size for 3D printing technologies, including two-photon printing (2PP) and two-step absorption ((1 + 1)PA). Diagonal lines mark constant volume rates, and an updated version is available at 3dprintingspeed.com. (e) From delivered light to finished parts: the bound or definition behind each throughput metric, the information needed for a fair comparison and an example from the reviewed studies, together with the quality metrics that should accompany every rate. Symbols: $R_{vox}$ voxel rate, $\eta$ delivery efficiency, $P_{avg}$ average power, $E_{vox}$ energy per voxel (Equation (8)), $N$ number of foci, $f$ repetition rate, $M$ pulses per voxel, $v$ scan speed, $p$ voxel pitch, $n_{spot}$ spots per frame, $f_{frame}$ frame rate, $Q$ volume rate, $V_{vox}$ voxel volume, $\Phi$ fill factor, $A$ field area, $\Delta z$ layer thickness, $f_{layer}$ layer rate, $D$ fraction of time spent writing, $Q_{sus}$ sustained volume rate and $V_{part}$ part volume. (a,b) Adapted under the terms of the CC BY 4.0 license [15]. Copyright 2023, The Authors, published by Springer Nature. (c) Adapted under the terms of the CC BY 4.0 license [5]. Copyright 2026, The Authors, published by Springer Nature. (d) Adapted under the terms of the CC BY 4.0 license [23], updating the chart first published by Hahn et al. [22]. Copyright 2024, The Authors, published by Light: Advanced Manufacturing. Original panel labels were replaced. (e) Original schematic prepared for this review.

Peak and sustained rates also differ. Peak rates are usually measured over a single field during continuous exposure, whereas sustained rates include stage moves, stitching, layer changes, data transfer and repeated exposures for dose control. Where only peak values are reported, the time to print a finished part can be much longer than the voxel rate suggests.

A comparable report of parallel TPL throughput should therefore include at least the following (Figure 4e):

- the voxel definition and size, including whether each focal spot counts as one voxel
- peak and sustained voxel rates, including stage moves and stitching
- the volume rate and fill factor for a specified test geometry
- resolution metrics, namely the finest feature width and the smallest open pore
- the field size, stitching strategy and overlap between fields
- laser parameters, namely wavelength, pulse duration, pulse energy, repetition rate and power per focus
- the energy per voxel (average laser power divided by voxel rate) and, where the delivery efficiency is known, the energy reaching the resist ($E_{vox}$, Equation (8)).

The 7 × 7 printer of Kiefer et al. shows that most of these items can be reported in one paper. It specified the voxel definition (a two-photon Sparrow criterion giving about 200 nm × 500 nm), peak rates under two conventions, the laser parameters, the power per focus, the focus spacing and the measured power variation across the array [23]. What remains largely unreported, there and elsewhere, is the sustained rate for a finished part, although the two-day print time of their $4.62 \times 4.62 \times 10.26$ mm$^3$ metamaterial is a useful metric in its own right [23]. Similar standardization has begun for the mechanical testing of two-photon-polymerized materials [47].

The energy per voxel deserves particular attention. Under Equation (8), it sets the maximum voxel rate that a given laser can sustain, and it follows from parameters that many papers already report. The 7 × 7 printer used about 26 nJ of laser energy per voxel (3.7 W at $1.4 \times 10^8$ voxels s$^{-1}$) [23], whereas the 2000-focus holographic printer used about 2 μJ (4 W at $2 \times 10^6$ voxels

s$^{-1}$) [15]. This roughly 80-fold difference, which reflects, among other factors, the low optical efficiency of binary holograms (Section 5), is invisible in rate and resolution alone.

## 4 Fixed Multi-Focus Arrays

Splitting one beam into a fixed grid of foci is the simplest way to raise the voxel rate. The approach is reliable and now reaches about $10^8$ voxels s$^{-1}$, but because every focus follows the same path, the array prints copies of one pattern.

Two kinds of element generate the array (Figure 5a). In a microlens array (MLA), each lenslet forms a focus, whereas in a diffractive optical element (DOE), a phase pattern distributes the beam among diffraction orders that form a grid of spots in the focal plane. All foci share one scanning system and cannot be controlled individually, which generally limits these methods to periodic or replicated structures [15].

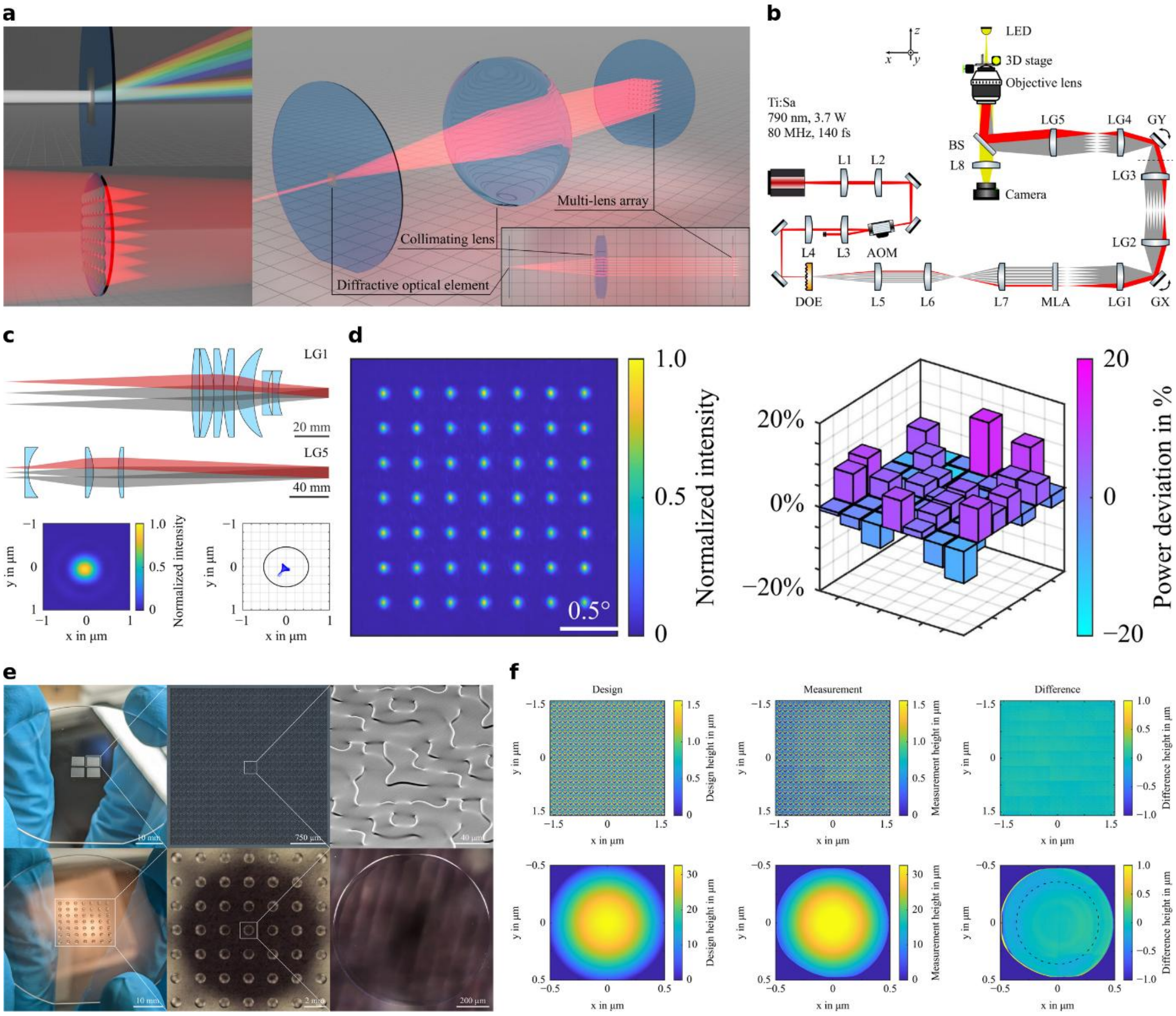

**Figure 5.** Fixed multi-focus arrays. (a) Beam splitting. A DOE disperses a broadband beam into wavelength-dependent diffraction orders (top left), an MLA focuses part of a Gaussian beam into foci

while half of the power passes unused (bottom left), and a DOE combined with a collimating lens and an MLA forms a 7 × 7 focus array (right). (b) Setup of the 7 × 7 printer. An acousto-optic modulator (AOM) controls the power of a Ti:sapphire laser (790 nm, 3.7 W, 80 MHz, 140 fs), the DOE splits the beam onto the MLA, and galvanometer mirrors (GX, GY) scan the array, which lens groups LG1–LG5 relay to the objective (L1–L8: lenses, BS: beam splitter). (c) Lens groups LG1 and LG5 with the outermost beam in red (top), and the simulated point-spread function and spot diagram of the outermost focus, with all rays inside the Airy ellipse (bottom). (d) Measured normalized intensity of the 49 foci (left) and deviation of each from the mean power (right). (e) The printed beam-splitting optics: a glass wafer carrying four DOEs, with micrographs of one DOE (top), and a wafer carrying the MLA, with micrographs of the array and one lens (bottom). (f) Height maps of the DOE (top) and one MLA lens (bottom), showing the design, the measurement after pre-compensation and their difference, with mean deviations of 71 and 142 nm. (a–f) Adapted under the terms of the CC BY 4.0 license [23]. Copyright 2024, The Authors, published by Light: Advanced Manufacturing. Original panel labels were replaced.

The limiting case of fixed parallel exposure is multi-beam interference, which writes an entire periodic lattice at once. Campbell et al. fabricated photonic crystals by four-beam holographic lithography [48], and Kondo et al. interfered femtosecond pulses from a diffractive beam splitter to write one-, two- and three-dimensional periodic microstructures [49]. Interference offers the highest parallelism but only periodic structures, which motivates scanned focus arrays that can write arbitrary unit cells.

Microlens arrays were among the first focus-array approaches, used in 2005 for multiple-spot femtosecond processing [50] and periodic microstructures [51]. They established the central trade-off of fixed arrays: much higher throughput for repeated patterns at the cost of design freedom.

DOEs have since become more common. Arnoux et al. designed binary-phase spot-array DOEs with a modified iterative Fourier-transform algorithm, producing inter-spot diffraction angles from 1° to 0.01° that set the focus spacing in the resist [7]. Maibohm et al. fabricated a 3 × 3 DOE in photoresist on glass and matched the 50 µm beamlet pitch to the unit-cell size of their scaffolds. About 70% of the 148 mW input reached the beamlets, roughly 11.5 mW per focus, and with a 0.75 NA objective the fabrication time fell ninefold, to 37 min for 81 unit structures across 450 × 450 $\mu m^2$ [8].

Stitching remained a limitation. Stage moves of more than 150 µm between copies opened gaps in thin walls until the unit cell was redesigned so that neighbouring copies overlapped within supporting towers, which reduced mismatches to below 100 nm [8]. Array geometry and part design must therefore be optimized together.

Combining a DOE with fast galvanometric scanning increased throughput further. Hahn et al. scanned a 3 × 3 array at about 0.5 m $s^{-1}$, approaching $10^7$ voxels $s^{-1}$ with sub-micrometre voxels, and printed a chiral mechanical metamaterial with more than $10^5$ unit cells and $3 \times 10^{11}$ voxels [22].

Kiefer et al. extended this approach to a 7 × 7 array [23]. Doubling the focus velocity to 1 m $s^{-1}$ with a 1.4 NA objective near the diffraction limit raised the throughput tenfold, to about $10^8$ voxels $s^{-1}$. A 790 nm laser delivered 3.7 W in 140 fs pulses at 80 MHz, giving each of the 49 foci, spaced 60 µm apart, about 19.5 mW (Figure 5b). Custom relay lenses kept even the outermost focus diffraction-limited (Figure 5c), and the power varied by ±13% across the array (Figure 5d) [23]. The DOE and the matching lens array were themselves printed by two-photon grayscale lithography, with mean deviations from the design of 71 and 142 nm after pre-compensation (Figure 5e,f) [23]. TPL can thus print the micro-optics that parallel TPL itself relies on.

These demonstrations reached large part volumes by TPL standards. A 4.62 × 4.62 × 10.26 $mm^3$ chiral metamaterial with about one million unit cells and more than $1.7 \times 10^{12}$ voxels took two days to print, and the same printer produced about 7 million carrier microparticles, each about 50 µm across, for potential pharmaceutical use [23].

Metasurfaces offer another route to fixed foci. X. Wang et al. encoded seven foci in a shared-aperture silicon nitride metalens, which they described as the first use of a metasurface in 3D nanolithography. The metalens achieved 95.21% intensity uniformity and a sevenfold gain in throughput while retaining a high numerical aperture with a longer working distance than a metalens array [52].

Fixed arrays are also expanding into three dimensions. In a 2026 preprint, Le Deun et al. used a static DOE to generate two spot arrays in planes 1.8 µm apart. With 29 spots, they printed four-layer woodpiles at 1 $mm^2$ in 90 s, reducing the number of passes without adding moving parts [53].

The focus spacing sets the separation between copies and the strength of their interaction [7]. In the 7 × 7 printer, the outermost foci are 360 µm apart and must fit within the corrected field of the objective while leaving scan range [23], so larger arrays need wider-field optics or, as in metalens arrays, a separate lens for each focus [46].

Because the power of each focus is difficult to adjust independently [5], uniformity depends on element design and alignment [8], and as the spacing decreases, optical overlap and radical diffusion cause crosstalk (Section 10) [7]. Within these constraints, fixed arrays remain the established approach for printing identical parts and periodic structures rapidly.

## 5 Reconfigurable Multi-Focus Writing

Programmable modulators remove the restriction to one pattern. A computer-generated hologram on an SLM lets each focus be positioned, attenuated or switched off independently, so the same optics can write different objects simultaneously or distribute one aperiodic object over many foci. This flexibility costs modulator speed, damage threshold, optical efficiency and focus uniformity.

Two modulator types dominate. Liquid-crystal SLMs modulate the phase and can direct nearly all incident light into the foci, but they refresh slowly. DMDs switch much faster and tolerate higher power but provide only binary amplitude control [5], and binary holograms direct only a fraction of the light into the useful diffraction order, trading optical efficiency for speed.

Multi-focus holograms are usually computed iteratively. The Gerchberg–Saxton algorithm propagates the field between the modulator and focal planes, enforcing the modulator constraint in one plane and the target foci in the other, and its weighted variant improves uniformity. DMDs instead use binary Lee holograms, which encode the phase in the fringe positions of a binary grating. Fixed DOEs use related algorithms computed once, before fabrication [7], and an SLM can also correct aberrations by imposing a compensating phase [34].

Hayasaki et al. first demonstrated holographic parallel femtosecond processing with a liquid-crystal SLM [54], and Jenness et al. used SLM phase holograms for 3D parallel micropatterning [55]. Obata et al. applied individually controlled foci to two-photon polymerization, fabricating differently shaped 3D objects simultaneously rather than copies [56].

Gittard et al. used 16 foci to write arrays of micro-Venus figures and four foci for tissue-engineering scaffolds and microneedle arrays, about four times faster than single-beam writing at the same speed and energy [57]. Zandrini et al. fabricated Nichoid scaffolds for stem-cell culture with up to six foci. The total power rose from 21 mW for one focus to 260 mW for six, and cell viability was comparable to that on single-focus scaffolds [58]. L. Yang et al. replaced repeated scanning with predefined multi-focus patterns to fabricate microlens arrays and 3D photonic structures in parallel [59].

When holograms reposition the foci, throughput depends on how quickly new patterns can be computed and displayed. Vizsnyiczai et al. calculated holograms on a graphics processor at the 60 Hz refresh rate of their SLM and coordinated five foci, needing 549 holograms to place 2745 voxels in one microstructure [60]. At 60 Hz, five steered foci can place at most 300 voxels $s^{-1}$.

Liquid-crystal SLMs also have low damage thresholds, which limits their use with high-power femtosecond beams [5].

DMDs update much faster. Geng et al. displayed binary Lee holograms on a DMD to generate and steer one to tens of foci at 22.7 kHz, more than 300 times the SLM rate above (Figure 6a). With one to three foci, they printed 3D trusses and woodpiles at about 500 nm resolution (Figure 6b), and random-access placement suited structures that are hard to raster-scan, including overhangs (Figure 6c) [61]. Combined with aberration correction, DMD holography printed large areas without stitching [62].

Ouyang et al. scaled DMD holography to 2000 independently programmable foci, using the weighted Gerchberg–Saxton algorithm to set the amplitude, phase and position of each (Figure 6d–f) [15]. With 100 fs pulses at 800 nm and 1 kHz (4 W), single-pulse polymerization yielded 90 nm resolution at $2 \times 10^6$ voxels $s^{-1}$. A custom photoinitiator with a TPA cross-section of about 800 GM was required, and for lattices with fill factors of 1–12%, 2000 foci corresponded to 4.5–54.0 $mm^3$ $h^{-1}$ [15]. The group later reached a focus uniformity above 99.9% and, with randomized 25-point exposures, wrote a 1 $mm^2$ diffractive element with four million optical neurons at 500 nm pixel resolution in 15 min [63].

Holography can also be combined with fast scanning, in which a scanner moves the focus pattern of each hologram through the resist. Zhang et al. combined galvanometric scanning with more than 400 SLM-generated foci to reach $1.49 \times 10^8$ voxels $s^{-1}$ and identified generating many uniform foci as the main challenge [64].

A hybrid approach keeps a fixed focusing element but makes its illumination programmable. S. Yang et al. combined a DMD with an MLA so that thousands of foci could be switched and graded individually, printing uniform dot arrays with features below 200 nm [65]. Gu et al. scaled this concept to more than 120,000 foci, formed by a 12 $cm^2$ metalens array and individually switched and graded by an SLM, and reached rates above $10^8$ voxels $s^{-1}$ [46]. By compressing sub-blocks of a structure into shared patterns and switching spots as the stage scanned, they printed fully aperiodic geometries with features down to 113 nm, including chainmail, octet and Kelvin lattices and terahertz metamaterials, as well as more than 50 million microparticles per day [46].

Hologram quality remains an active topic. K. Liu et al. attributed speckle and focal non-uniformity to the Gerchberg–Saxton algorithm and replaced it with model-driven deep learning, reaching $1.2 \times 10^5$ voxels $s^{-1}$ at 120 nm resolution with up to 2000 voxels per exposure [66]. C.

Wang et al. generated volumetric focus arrays with greater than 90% consistency in all three dimensions to print spiral arrays with circular dichroism [67], and J. Li et al. used SLM beam splitting for nanoscale lithography of photonic crystals in yttrium aluminium garnet, with gaps as narrow as 261 nm (Figure 6g,h) [68].

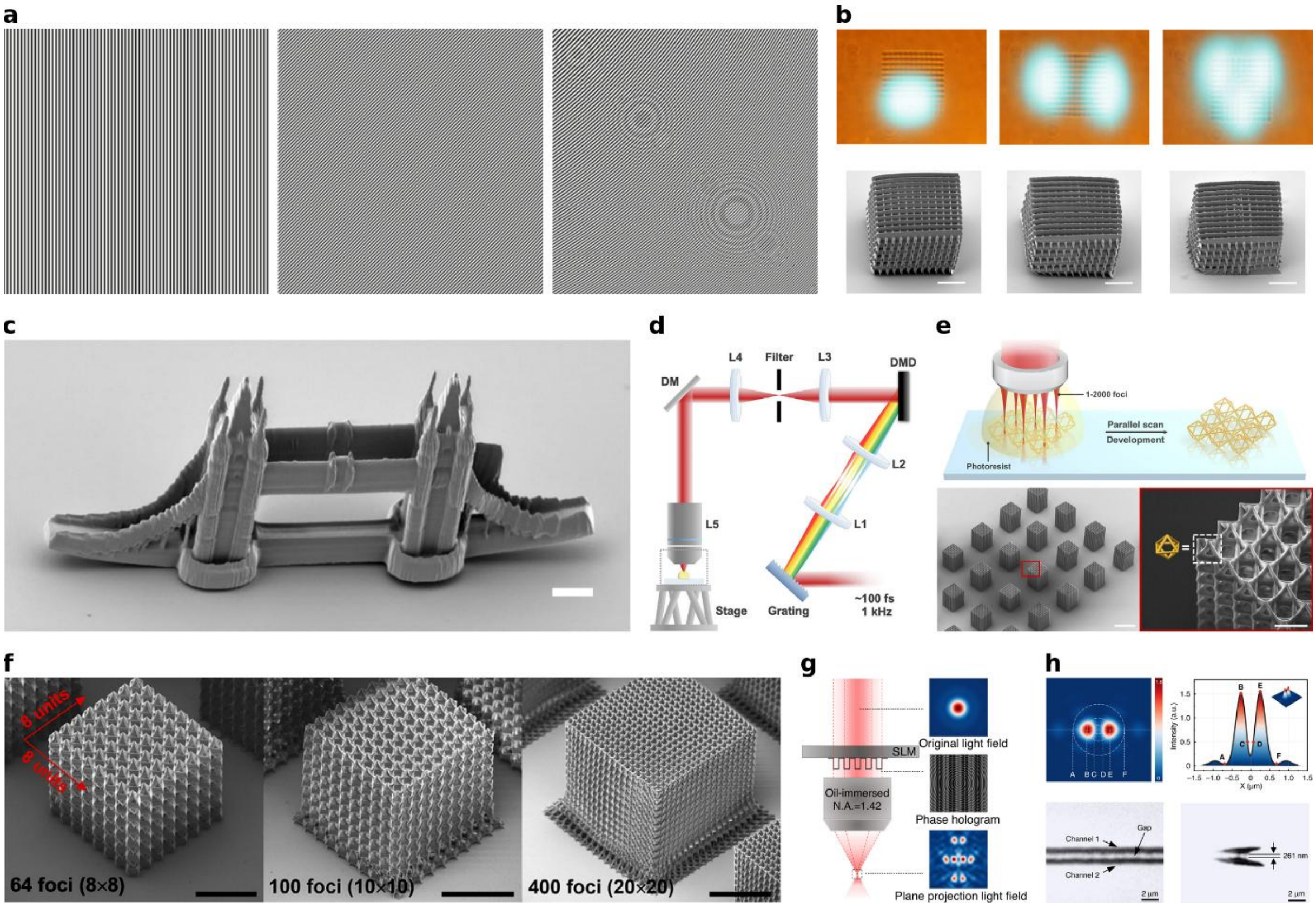

**Figure 6.** Reconfigurable holography. (a) Binary DMD holograms for scanning along *x* (left), in the *x*–*y* plane (middle) and in 3D random access (right). (b) Woodpile printing with one, two and three foci (top) and electron micrographs of the structures (bottom). (c) Microscale London Bridge printed by random-access DMD scanning. Scale bars in (b,c), 10 μm. (d) Setup of a DMD-based holographic printer driven by ≈100 fs pulses at 1 kHz (DM: dichroic mirror, L1–L4: lenses, L5: objective). (e) Parallel writing of octahedral truss lattices with 1–2000 foci (top) and electron micrographs of truss blocks printed with 64 foci (bottom, scale bars 100 and 10 μm). (f) Metastructures printed with 64, 100 and 400 foci (scale bars, 50 μm). (g) Multi-beam light field from a phase hologram on an SLM, focused through a 1.42 NA objective. (h) Double-beam writing in yttrium aluminium garnet: simulated focal intensity (top) and micrographs of two channels separated by a 261 nm gap (bottom, scale bars 2 μm). (a–c) Adapted under the terms of the CC BY 4.0 license [61]. Copyright 2019, The Authors, published by Springer Nature. (d–f) Adapted under the terms of the CC BY 4.0 license [15]. Copyright 2023, The Authors, published by Springer Nature. (g,h) Adapted under the terms of the CC BY 4.0 license [68]. Copyright 2023, The Authors, published by Springer Nature. Original panel labels were replaced.

Reconfigurable writing still has clear limits. Holographic foci cannot be packed arbitrarily closely (Section 2.4) [5], and each must reach about 1 TW cm$^{-2}$, so adding foci raises the required laser power [15]. Liquid-crystal SLMs are efficient but slow and damage-prone, whereas DMDs are fast but waste much of the light. The strength of holography is its geometric flexibility, which makes it, alone or combined with fixed focusing arrays, the main route to printing aperiodic structures at high throughput.

## 6 Shaped Beams and Single-Exposure Writing

Multi-focus methods still scan each focus. An alternative is to reshape a single beam so that its focal region matches the target structure, or one of its building blocks, and a single exposure then polymerizes an entire tube, ring or cage. Balena et al. identified this approach as one of the two main routes to faster two-photon writing [26].

Shaped exposures emerged almost as early as multi-focus writing. Bhuian et al. wrote annuli up to 50 µm in diameter with an axicon-profiled femtosecond beam [69], and Phelan et al. used internal conical diffraction to write sub-micrometre rings and, with the associated Bessel beam, pillars and walls with height-to-width ratios of five to eight [70]. SLMs later made such shaping programmable.

Bessel beams are the classic example. Their transverse profile is nearly invariant over long axial distances, and higher-order beams have ring-shaped cross-sections, so one exposure can define an entire tube. L. Yang et al. used SLM-generated Bessel beams to polymerize complete microcylinders, 15.5–55 µm in diameter and 11.2 µm tall, in single exposures of 5 or 10 s, avoiding the cubic scaling of serial writing time [71]. Scanning such a beam produced microtube arrays two orders of magnitude faster than point-by-point writing, and these arrays served to capture, transfer and release cells [72].

Other propagation-invariant beams write other shapes. C. Wang et al. used Mathieu beams to print microcages for trapping micro-objects, and a single-layer cage took 0.36 s instead of 54 s by point scanning, while stacked exposures produced three-layer cages about 54 µm tall [73].

Abruptly autofocusing beams, first used for multiscale photopolymerization by Manousidaki et al. [74], propagate along a ring that contracts nonlinearly and produce a sharp rise in intensity near a chosen depth [75]. Jia et al. designed abruptly autofocusing vortex beams whose rings of 15.6 and 10.4 µm contracted to 4.4 and 2.6 µm at the focal plane (Figure 7a–c), and they mapped the printable window in laser power and exposure time [75]. The walls were much thicker than the 2.9 µm main lobe because the side lobes exceeded the threshold (Figure 7d,e), as estimated in Section 2.4. A 10 × 10 array of microtraps took about 3 min to print and captured microspheres with a success rate above 90% (Figure 7f) [75].

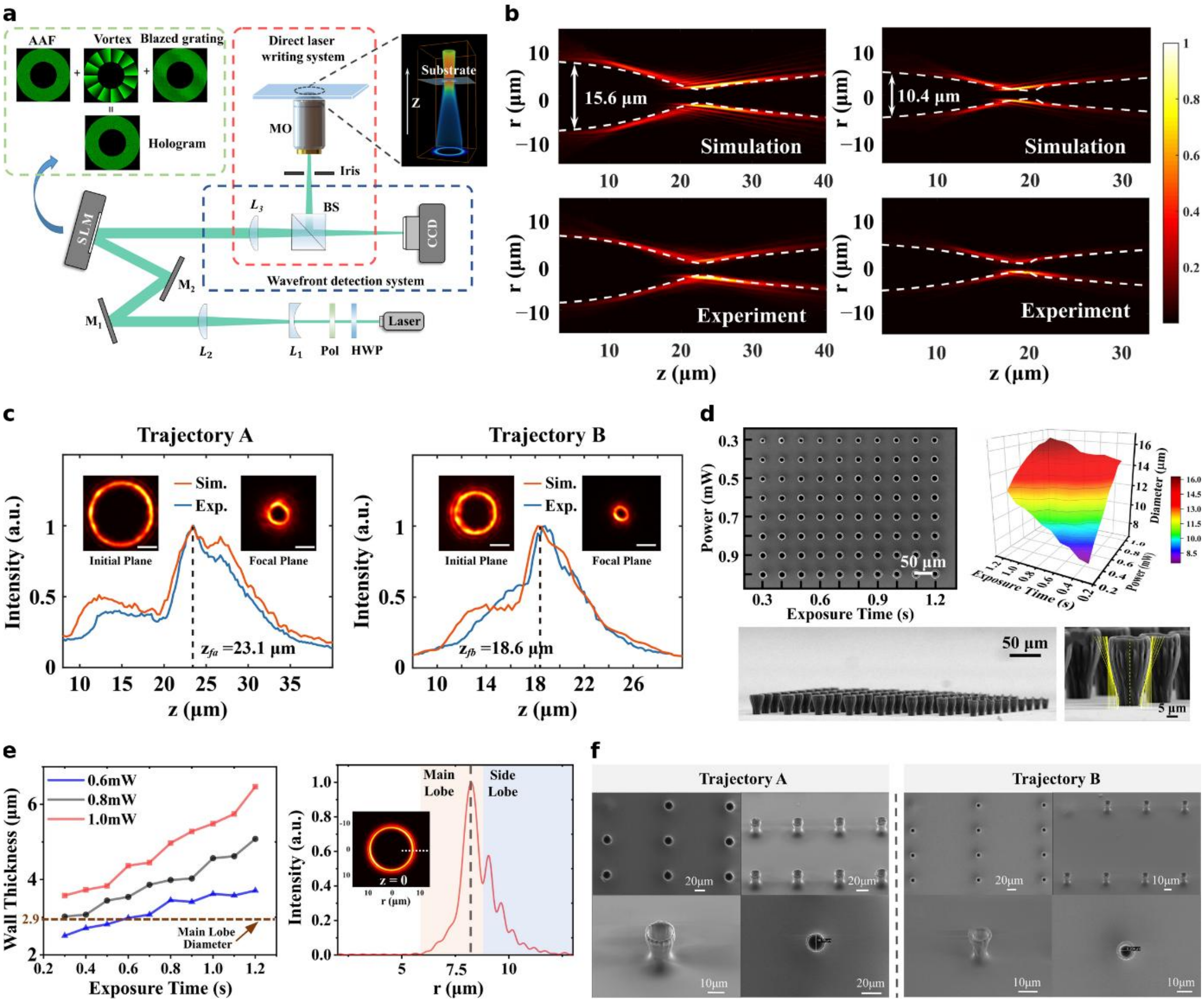

**Figure 7.** Shaped beams. (a) Setup for abruptly autofocusing vortex (AAFV) beams. A phase-only SLM displays a hologram that combines an abruptly autofocusing phase, a vortex phase and a blazed grating, and a wavefront detection system records the beam (HWP: half-wave plate, Pol: polarizer, M1 and M2: mirrors, L1–L3: lenses, BS: beam splitter, MO: microscope objective). (b) Simulated (top) and measured (bottom) intensity of AAFV beams in the *r*–*z* plane for trajectories A and B, with initial ring diameters of 15.6 and 10.4 µm. Dashed lines show the designed caustics. (c) Simulated (orange) and measured (blue) peak ring intensity along *z*, with insets showing the initial and focal beam profiles (scale bars, 5 µm). (d) Microtraps printed at 0.3–1.0 mW and 0.3–1.2 s, their inner-opening diameter, a printed array and one trap contour compared with the design. (e) Wall thickness versus exposure time at 0.6–1.0 mW, with the 2.9 µm main lobe indicated (left), and the focal intensity profile with its side lobes (right). (f) Electron micrographs of microtrap arrays printed with trajectories A and B. (a–f) Adapted under the terms of the CC BY 4.0 license [75]. Copyright 2023, The Authors, published by MDPI. Original panel labels were replaced.

Holography can also project a whole 2D pattern into the focal plane in one exposure. L. Yang et al. showed how the focal intensity distribution, the pattern design and the dose determine the profile and surface roughness of such projected microstructures [76]. Section 7 describes how temporal focusing extends projection to 3D.

One-photon single-exposure holography provides a useful comparison. Lin, Baker and Menon projected a stable 3D intensity pattern from an inverse-designed phase mask through thick photoresist, fabricating 800 × 800 × 720 $\mu m^3$ structures in about 20 s with aspect ratios above 120:1 and 6 µm features [77]. Counted in 6 µm voxels, this volume corresponds to roughly $10^5$ voxels $s^{-1}$, a high volume rate but with far coarser features than two-photon writing. A related study reports more than $10^6$ addressable voxels in a single 7.5 s exposure, above $10^5$ voxels $s^{-1}$ [78].

Shaped beams trade geometric generality for speed. Each beam family produces characteristic forms, so the approach suits parts assembled from repeated functional units. The extended focus also complicates dose control, because the intensity varies along the structure, and most demonstrations remain proofs of principle, often for trapping cells or particles. Scanning a shaped focus [72] or replicating it with a multi-focus hologram, as in Bessel-beam arrays [79], multiplies the gains.

## 7 Projection with Temporal Focusing

Projection TPL images a pattern, typically displayed on a DMD, into the resist and cures an entire layer at once. Each bright pixel or pixel group acts as a focal spot, so the number of simultaneous spots is set by the modulator rather than by beam splitting.

A conventionally focused wide-field image would cure a thick slab, because broad illumination barely varies along the optical axis. Temporal focusing restores axial confinement. A grating, or the DMD itself, disperses the spectral components of each pulse, and the objective recombines them only at the focal plane, where the pulse is therefore shortest and most intense (Figure 8a–c). The principle comes from scanless two-photon microscopy [80] and was demonstrated independently in 2005 by Oron, Tal and Silberberg, who acquired depth-resolved images without scanning [81], and by Zhu et al., who described simultaneous spatial and temporal focusing [82].

Kim and So applied temporal focusing to 3D lithography in 2010, fabricating microstructures in SU-8 with an axial resolution consistent with theory [83], and Li et al. combined it with patterned excitation to accelerate the fabrication of freeform polymer microstructures [84].

Saha et al. turned the concept into a scalable printer that wrote arbitrary 3D structures layer by layer from digital masks, raising throughput by up to three orders of magnitude without sacrificing sub-micrometre resolution [19]. With 35 fs pulses of about 1 mJ at 1 kHz, each DMD pixel mapped to about 151 nm, and up to about one million pixels were projected per exposure. The printer wrote nanowires 130–140 nm wide and built a 2.20 × 2.20 × 0.25 $mm^3$ cuboid in 8 min 20 s instead of several hours [19].

Projection is demanding on the laser, because the entire field must reach about 1 TW $cm^{-2}$ within the same pulse [19], so the usable field grows with the pulse energy. Projection also uses light differently from holography. A phase hologram moves power from dark regions into the foci, whereas a DMD discards the light that falls on switched-off mirrors. Projection is therefore more efficient for dense layers and holography for sparse patterns, consistent with the fill-factor argument of Ouyang et al. [15].

Somers et al. extended projection to continuous printing (Figure 8a) [44]. Their DMD both patterned the beam and, because its mirrors tilt along their diagonals, acted as the grating for temporal focusing (Figure 8b), while the stage moved axially at 100 µm $s^{-1}$ as the layers streamed. With 65 fs pulses at 5 kHz and a 1.49 NA objective, each mirror mapped to about 51 nm, the field measured 43.81 × 24.66 $\mu m^2$ and line widths were below 200 nm. A microlens took less than 120 ms to print, and a 1 $mm^3$ lattice of 42 × 42 × 42 unit cells took 2.3 h (Figure 8d) [44]. The authors noted that curing large fields requires more power than femtosecond oscillators provide [44].

Dose control was a major limitation of early projection systems. Dense regions over-polymerize and sparse ones under-cure, because light and reactive species from neighbouring spots add to the local exposure [5, 7]. Projecting images sequentially or varying the pulse count per layer reduces throughput and prevents single-pulse layers [5]. Kim, Pingali and Saha separated the optical and chemical origins of these proximity effects, and their interspersing projection printed features below 300 nm and pores below 700 nm at more than 0.5 $mm^2$ $s^{-1}$ per layer, up to 50 times faster than point scanning [45]. A later study of more than 10,000 parameter combinations found no physics-based lower limit to feature size in a well-calibrated projection system. In practice, sub-100 nm features are limited by the sensitivity of polymer conversion to the processing conditions [85].

Kim and Saha then turned to dose control with grayscale projection [5]. Each mirror of their 1920 × 1080 DMD mapped to 114 nm in the resist, and each focal spot of about 785 nm was defined by a 7 × 7-pixel motif (Figure 8e). Because the motif is below the diffraction limit, switching off pixels lowers the spot intensity without proportionally shrinking the spot. The 32 motifs spanned intensities from 100% to 10% and varied the printed line width from 850 to 275 nm (Figure 8f). With 35 fs pulses at 5 kHz and the grayscale encoded spatially rather than in the exposure time, a single pulse wrote each layer, giving 5000 layers per second during continuous axial motion. With 15,836 tunable spots per image, the peak rate was $7.9 \times 10^7$ voxels $s^{-1}$ with each spot counted as a voxel, or $1.7 \times 10^9$ voxels $s^{-1}$ with voxels defined by the finest feature (Section 3). The method also produced nanowires as thin as 55 nm and reached about 215 $mm^3$ $h^{-1}$ (Figure 8g). For nanoporous structures, it was 270 times faster than interspersing projection and 3000 times faster than high-speed serial TPL, and heating of the DMD limited the usable laser power [5].

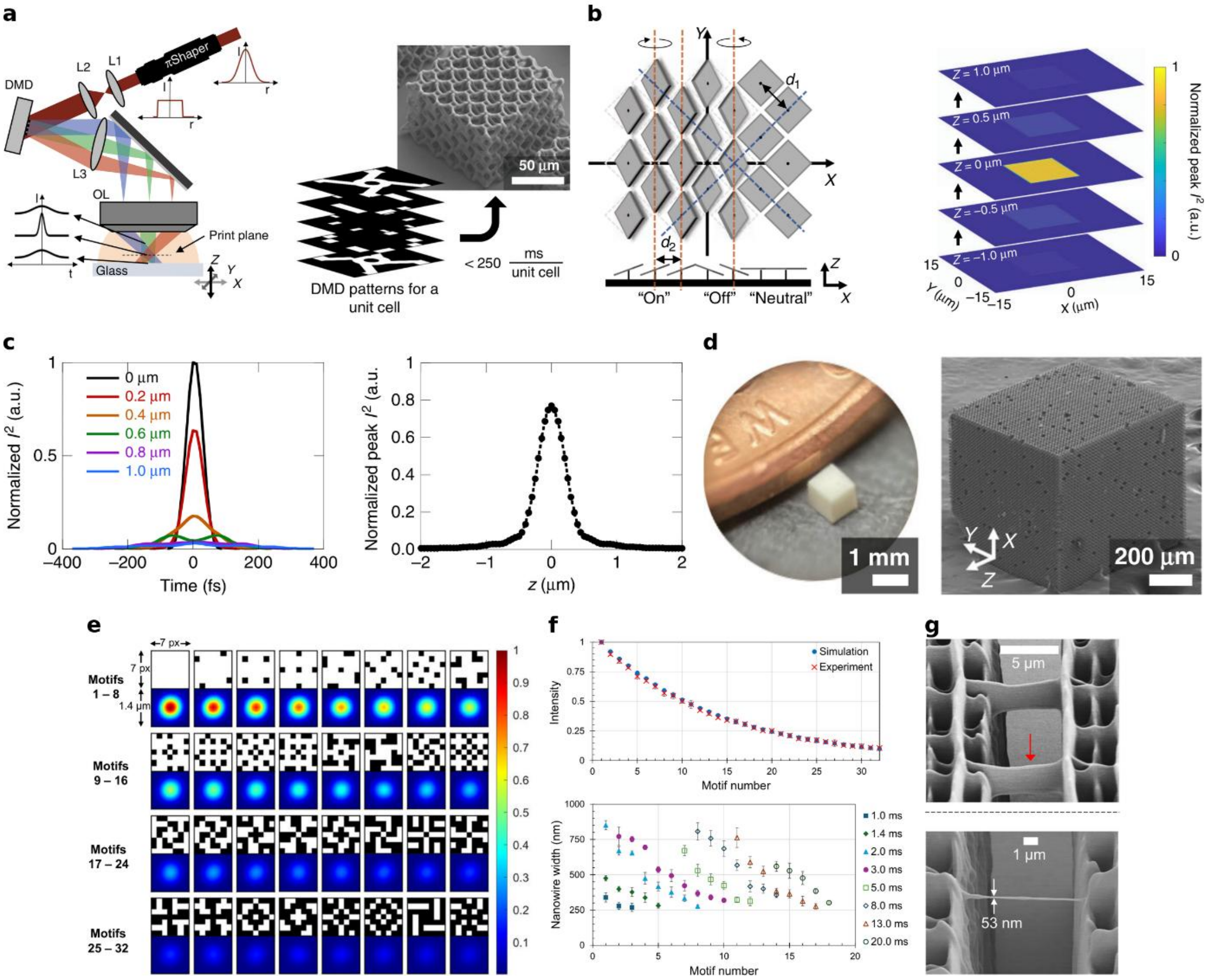

**Figure 8.** Projection with temporal focusing. (a) Continuous projection. A flat-top beam illuminates the DMD, whose pattern the objective lens (OL) images onto the print plane, where temporal focusing compresses the pulse (left). A stack of DMD patterns prints a unit cell in less than 250 ms, here forming a 5 × 5 × 3 lattice (right). (b) The DMD as a grating. The mirror grid forms a two-dimensional grating of period $d_1$ and the mirror tilt a one-dimensional grating of period $d_2$ (left), and the simulated peak $I^2$ is shown in planes from $z = -1.0$ to 1.0 μm for a projected rectangle (right). (c) Simulated pulse profiles at different distances from the print plane (left) and peak $I^2$ along the axis (right). (d) A 42 × 42 × 42 unit-cell structure beside a one-cent coin (left) and its electron micrograph (right). (e) The 32 binary 7 × 7-pixel motifs used for grayscale projection and their simulated intensity profiles. (f) Simulated and measured relative peak intensity of each motif (top) and widths of nanowires printed with each motif at exposure times of 1.0–20.0 ms (bottom). (g) Suspended wires printed with fully illuminated motifs (top) and a 53 nm nanowire printed with grayscale motifs (bottom). (a–d) Adapted under the terms of the CC

BY 4.0 license [44]. Copyright 2021, The Authors, published by Springer Nature. (e–g) Adapted under the terms of the CC BY 4.0 license [5]. Copyright 2026, The Authors, published by Springer Nature. Original panel labels were replaced.

Projection is also moving beyond kilohertz amplifiers. In a 2026 preprint, Papamakarios et al. combined holographic printing with spatiotemporal beam shaping around a 500 kHz, 515 nm laser, choosing the photoinitiator to suppress thermal crosstalk, and exceeded $2.3 \times 10^6$ voxels $s^{-1}$ at resolutions below 400 nm [86]. In another preprint, Zhong et al. temporally focused pulses into a DMD-programmed line with pixel-level grayscale at more than 10 kHz. Continuous line scanning with graded exposure at the seams eliminated stitching defects in centimetre-scale structures with 75 nm lateral features, and pattern streaming at a bandwidth an order of magnitude higher than in previous TPL systems showed that data transfer could become a bottleneck [87].

Area projection exposes many spots per pulse but needs high pulse energy and stitching between fields. Line projection needs less energy per pulse and supports continuous scanning, but it requires more frequent pattern updates. The choice therefore shifts the bottleneck between the laser, the modulator and the data link.

Projection nevertheless has intrinsic constraints. Because the whole field must reach threshold simultaneously, the pulse energy limits the field size and loads the modulator heavily [5, 44]. Proximity effects also extend across the image, large parts still require stitching (Section 10), and layer-by-layer projection cannot build large overhangs without supports [15]. In return, projection writes an entire layer per pulse and, with grayscale motifs, controls the dose spot by spot.

## 8 Routes Beyond Two-Photon Absorption

TPA is not the only route to confined curing. Any excitation whose rate scales quadratically with intensity provides confinement, and some alternatives work at far lower intensities. These methods are not TPL in the strict sense, but they compete for the same applications and relax the power demands that constrain parallel writing.

Hahn et al. classify such processes, including upconversion, two-step absorption and triplet–triplet annihilation, as (1 + 1)-photon absorption. Excitation proceeds through a real intermediate state, which removes the need for intensities of about 1 TW $cm^{-2}$ and allows inexpensive continuous-wave lasers to be used [27].

Two-step absorption has progressed furthest as a printing method. Hahn et al. printed 3D nanostructures with about 100 µW from a 405 nm continuous-wave laser diode [88]. Messer et al. then built a shoebox-sized printer from a sub-milliwatt GaN laser diode, a miniature MEMS scanner, a 1.4 NA objective and microcontroller electronics. It achieved a lateral resolution of about 100 nm at about 1 mm $s^{-1}$ (Figure 9), whereas commercial printers are built around large, expensive ultrafast lasers [89].

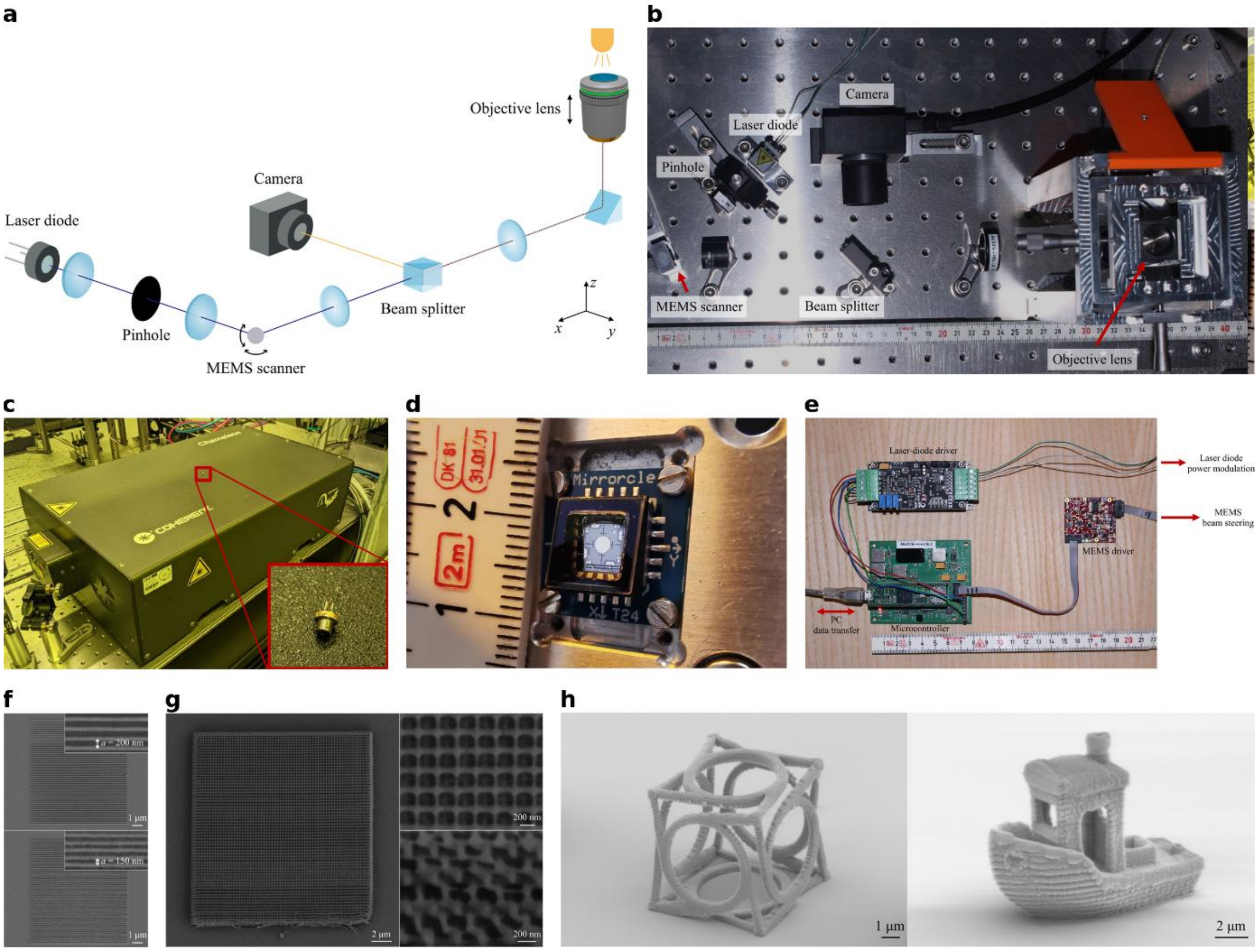

**Figure 9.** A shoebox-sized 3D laser nanoprinter based on two-step absorption. (a) Optical scheme. Light from a 405 nm laser diode passes a pinhole, is steered by a two-axis MEMS scanner and is focused into the photoresist by an objective moving along *z*, while a camera monitors printing. (b) The complete setup, with a ruler for scale. (c) A femtosecond oscillator of the type used in TPL, with the much smaller laser diode in the inset. (d) The MEMS scanner (15 × 15 $mm^2$ package, 1.6 mm mirror). (e) The microcontroller-based control unit for power modulation and beam steering. (f) Line gratings with periods of 200 nm (top) and 150 nm (bottom). (g) A woodpile with a 300 nm lateral period and 100 nm layer spacing, with close-ups. (h) A unit cell of a chiral mechanical metamaterial and a #3DBenchy boat, both printed at 1 mm $s^{-1}$. (a–h) Adapted under the terms of the CC BY 4.0 license [89]. Copyright 2024, The Authors, published by Light: Advanced Manufacturing. Original panel labels were replaced.

The 405 nm writing wavelength [88, 89] is about half that of Ti:sapphire lasers, which shrinks the diffraction-limited focus at the same numerical aperture. Whether this yields smaller features, however, depends on the resist and its intermediate state [27].

The low power per voxel makes these routes attractive for parallel writing. Rietz et al. combined a two-step-absorption resist with DMD-based binary holography to split, steer and grade the beam, reaching 61,500 voxels $s^{-1}$ with up to 31 continuous-wave foci [42]. Because the intermediate-state lifetime limits the single-focus scan speed in these resists, more foci are the direct route to higher throughput [42].

Light-sheet printing extends the concept to whole planes. Hahn et al. intersected images projected with 440 nm laser diodes with a 660 nm light sheet, so that curing occurred only where the colours overlapped. They reached $7 \times 10^6$ voxels $s^{-1}$ at a voxel volume of 0.55 $\mu m^3$, whereas linear projection had reached about $10^6$ voxels $s^{-1}$ only with voxels larger than 100 $\mu m^3$ [90]. Like projection TPL, light-sheet printing exposes whole planes and needs a second mechanism for axial confinement, in this case the intersection of two colours rather than temporal focusing.

Xolography, a linear relative, shows the volume rates that become attainable when resolution requirements are relaxed. Using photoswitchable initiators that polymerize only where two colours intersect, Regehly et al. reported about tenfold finer resolution than computed axial lithography [91] and volume rates four to five orders of magnitude above those of two-photon printing [92]. Somers et al. compare such volumetric methods with multiphoton printing [24].

At present, these approaches trade feature size or throughput for cheaper illumination. A 0.55 $\mu m^3$ light-sheet voxel is far coarser than the 55 nm wires of grayscale projection [5, 90], and 61,500 voxels $s^{-1}$ is far below the $\approx 10^8$ voxels $s^{-1}$ of femtosecond focus arrays [23, 42]. Their appeal lies in low cost and size and in the combination of low-power excitation with extensive parallelization. If resist design can relax the lifetime limit without compromising resolution, two-step absorption could lead to practical, low-cost parallel printers [27].

## 9 Enabling Hardware and Materials

Parallel optics pay off only if the rest of the printer keeps pace: the laser must supply every focus, modulators and scanners must update fast enough, the resist must cure at the resulting rate, and patterns must be computed and streamed without delay.

**Lasers.** The available power limits the number of foci that exceed threshold [8, 21]. DOE, DMD and SLM systems have used about 10–21 mW per focus [8, 58, 61], compared with about 100 µW for two-step absorption [88]. Optical efficiency matters too, as the 49 foci of the 7 × 7 printer received only about 1 W of the 3.7 W laser output [23].

The pulse format matters as much as the average power. With one pulse per voxel, the voxel rate equals the number of foci times the repetition rate, so 2000 foci at 1 kHz gave $2 \times 10^6$ voxels $s^{-1}$ [15]. Reaching $10^9$ voxels $s^{-1}$ would require a million foci per pulse at 1 kHz but only 2000 at 500 kHz. Amplifiers let thousands of foci or a whole projected field exceed threshold at once [5, 15, 44], with up to 2 μJ per focus before modulator losses, whereas the 80 MHz oscillator of the 7 × 7 printer supplied about 0.24 nJ per focus [23]. Single-pulse writing therefore pairs naturally with amplifiers, and oscillators suit fewer, rapidly scanned foci. Temporal focusing also benefits from the broad bandwidth of 35–60 fs pulses [5, 86].

**Modulators.** Liquid-crystal SLMs refresh at tens of hertz [60] and have relatively low damage thresholds. DMDs update binary patterns at kilohertz rates, up to 9523 Hz in grayscale projection [5], 22.7 kHz in holographic scanning [61] and more than 10 kHz with line illumination [87]. Their tilted mirrors also act as a grating that disperses femtosecond pulses, which projection exploits for temporal focusing [44] and holography must compensate [61].

**Static and adaptive optics.** DOEs split beams into well-defined arrays [7, 8], metalenses combine multiple foci with a high numerical aperture [52], and metalens arrays distribute foci over square centimetres with immersion compatibility and scalable fabrication [46]. An adaptive element can correct aberrations deep inside a material while shaping or splitting the beam [34], which maintains the focus quality across the field and through thick parts.

**Objectives, scanners and stages.** A high numerical aperture gives small, well-confined voxels but a small field of view, which limits the number of foci per field and increases stage moves. High-throughput arrays and projection systems use 1.25–1.4 NA [5, 23, 86], whereas voxels written at 0.75 NA were about 7 μm tall [8]. Galvanometers scan foci at 0.5–1 m $s^{-1}$ [22, 23], resonant scanners reach 66 m $s^{-1}$ [40], MEMS scanners make printers compact [89], and synchronized stages and galvanometers avoid stitching over tens of centimetres [39]. Projection instead moves the stage continuously while the patterns stream [5, 44]. Open hardware lowers the barrier to entry: an open-source holographic system with an SLM and a three-mirror scanhead writes voxels of about 250 nm × 1050 nm with three beams at 2000 voxels $s^{-1}$ [93].

**Resists.** Faster writing shortens each exposure, and the advantage of efficient initiators grows with scan speed. More soluble and efficient photoinitiators are therefore a priority for parallel printing [29], as the 800 GM initiator of the 2000-focus printer illustrates [15]. Large parts also need materials with ultralow shrinkage [17], because shrinkage distorts both the geometry and the joints between stitched fields. Chemistry can sharpen confinement, usually at a cost in sensitivity. Quenchers combined with photoinhibition achieved a 30 nm critical dimension [94], and ten-channel peripheral photoinhibition narrowed suspended lines from 47.1 to 20.2 nm at about $10^7$ voxels $s^{-1}$ [95].

**Data.** High-speed printers are also data-processing systems. Holograms must be computed as fast as they are displayed, whether on graphics processors [60], with weighted Gerchberg–Saxton algorithms [15] or by neural networks [66]. Pattern streaming can itself become the bottleneck [87], whereas encoding a whole structure in one overlaid pattern avoids per-focus data altogether [46].

## 10 Cross-Cutting Challenges

Parallel writing shifts the bottleneck from raw speed to control. Whatever the optics, the same challenges recur: dose non-uniformity, interactions between neighbouring voxels, heat, stitching, the trade-off between rate and resolution, and geometric constraints.

**Uniformity between foci.** As shown in Section 2.2, a 3% intensity difference near threshold can change the voxel diameter by almost 20%. Jie et al. observed exactly this between DMD-holographic foci. By adapting the voxel density, they improved the dimensional uniformity from 94.4% to 97.6% and printed a 6 × 5 array of supercritical lenses in 2 min [38]. Uniformity has become a standard metric, reported as 95.21% for a seven-focus metalens [52], more than 90% across volumetric arrays [67] and ±13% in power across the 49 foci of the 7 × 7 printer (Figure 5d) [23]. Generating many uniform foci is widely seen as a major difficulty [64, 65], compounded by speckle in iterative holograms [66].

**Proximity effects.** Arnoux et al. identified three regimes of spot spacing. Very closely spaced spots overlapped out of plane, which made parallel writing extremely difficult, intermediate spacings produced manageable diffusion-mediated crosstalk, and widely separated spots behaved independently [7]. Even within a 5 × 5 array with 2.22 μm spacing, curing thresholds differed between the centre and the corners. Yet by exposing only well-separated spots simultaneously, the authors could place periodic objects less than a micrometre apart [7]. Projection meets the same physics within each image [5], and reaction–diffusion models help separate the optical and chemical contributions [96]. Kim and Saha corrected the exposure spot by spot with proximity-compensation masks of 100%, 80% and 60% motifs. They also compensated a 3.3-fold intensity variation across a Gaussian field by assigning dimmer motifs to brighter regions, a method that transferred to another objective and to the laser ablation of gold films (Figure 10a,b) [5]. Grayscale exposure has also bent micropillars controllably (Figure 10d) [44] and tuned line widths across a metalens array [46].

**Heat.** At high repetition rates, heat accumulates and spreads between neighbouring voxels. Papamakarios et al. suppressed this thermal crosstalk at 500 kHz through their choice of photoinitiator [86], and heating of the DMD capped the usable laser power in grayscale projection [5].

**Stitching.** Parts larger than one field must be assembled from multiple exposures. Joints can be hidden inside supporting towers [8], formed as grayscale lap joints that reduce voxel broadening (Figure 10c) [5], or avoided by continuous line scanning with graded exposure [87] and by aberration-corrected DMD holography [62]. Gu et al. designed their centimetre-scale metalens array to overcome the limited field of view of conventional optics, a common source of throughput limits, proximity errors and stitching defects [46]. Low-shrinkage resists, in turn, keep the joints closed during development [17].

**Rate versus resolution.** Speed and fidelity remain in tension [5]. Ouyang et al. attributed their 90 nm resolution to single-pulse polymerization kinetics [15], but proximity effects close small pores in nanoporous structures (Figure 10e) [5]. Below 100 nm, the obstacle appears to be the sensitivity of polymer conversion to the processing conditions rather than a fundamental limit [85]. Photoinhibition reduces the feature size in parallel systems [95], although chemistries that sharpen features often reduce sensitivity [94].

**Geometric freedom.** Fixed arrays suit many identical parts [23], random-access holography suits sparse structures with 1–12% fill [15], and layer-by-layer projection suits dense layers, although overhangs need supports [15]. Lens arrays can write aperiodic structures by compressing sub-blocks into one pattern [46], but dense, aperiodic and overhanging geometries remain difficult for all approaches.

**Measurement and feedback.** Controlling thousands of foci requires feedback as well as calibration. Johnson et al. used Bayesian optimization with active learning to tune five pattern parameters of DMD-based projection printing. Within four iterations, the perimeter errors of circles, squares and triangles at three scales fell from about 0.5–1 µm to 116 nm or less (Figure 10f,g) [97]. The group later extended the approach to grayscale patterns guided by in situ quantitative phase imaging [98].

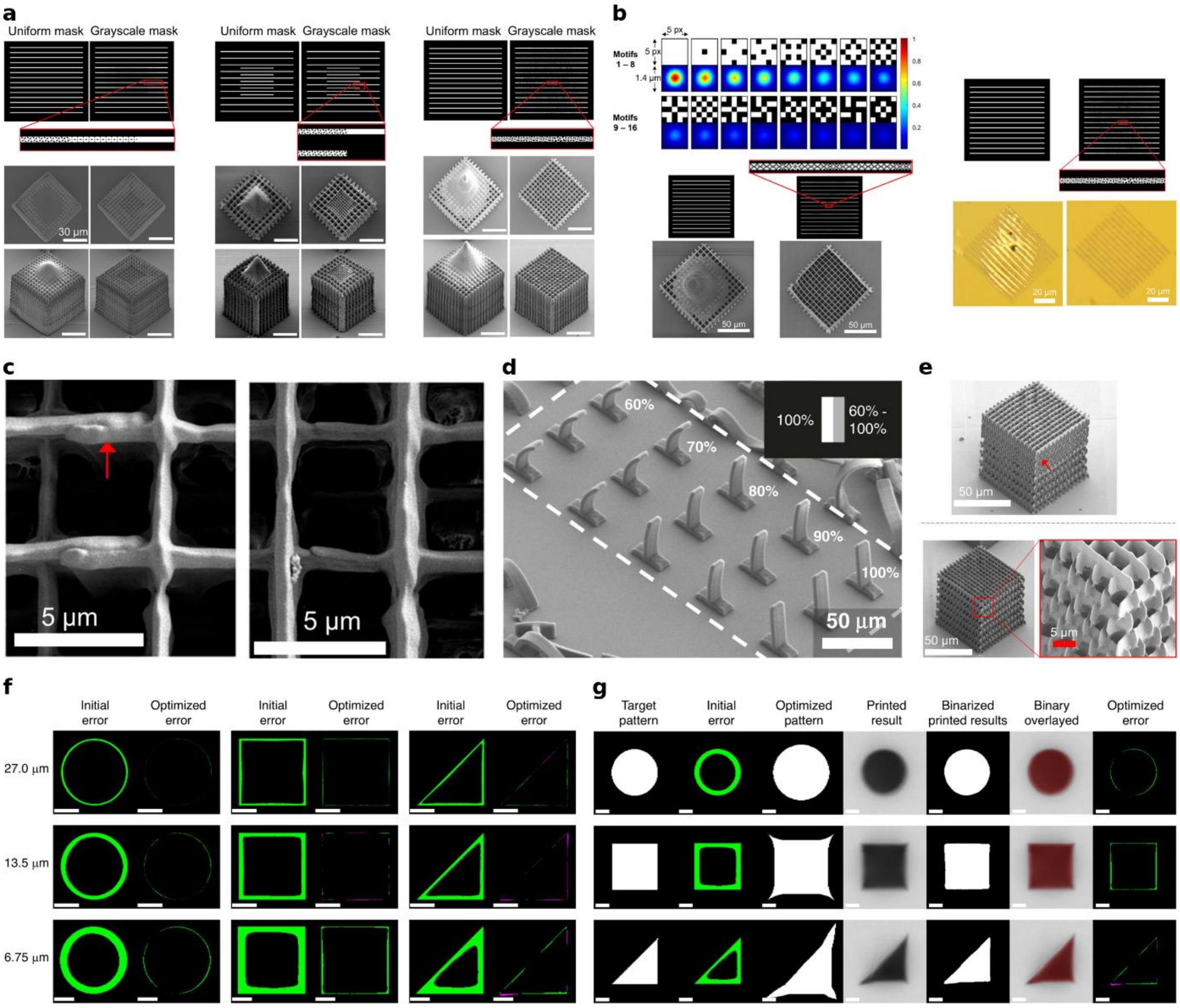

**Figure 10.** Controlling dose at scale. (a) Grayscale compensation for over-polymerization caused by proximity effects at woodpile centres (left, motifs at 100%, 80% and 60% intensity), higher feature density (middle) and the Gaussian beam profile (right, motifs from 100% to 31%). Each pair shows uniform masks and prints on the left and grayscale masks and prints on the right. Scale bars, 30 µm. (b) Transfer to other optics and processes: 5 × 5-pixel motifs for a 1.3 NA objective with their simulated intensity profiles (top left), polymer woodpiles printed with fully bright and Gaussian-compensated masks (bottom left), and line arrays ablated in a gold film with the same two masks, brighter regions indicating excessive removal (right). (c) Lap joints printed with overlapping fully bright motifs, showing a broadened joint (arrow, left), and with grayscale motifs in the overlap (right). (d) Micropillars printed by continuous projection with one half exposed at 100% and the other at 60–100% intensity (inset). Weaker crosslinking in the dimmer half increases shrinkage and bends each pillar. (e) Woodpiles printed with fully bright motifs, with pores blocked in the upper layers (top, arrows), and with grayscale masks, with open pores throughout (bottom). (f) Shape errors before (left of each pair) and after (right) Bayesian optimization of DMD patterns for circles, squares and triangles 27.0, 13.5 and 6.75 µm across. Green marks regions where the print is smaller than the target, and magenta where it is larger. Scale bars, 10, 5 and 2 µm (top to bottom). (g) Optimization steps for 6.75 µm shapes, from the target pattern and initial error to the optimized pattern, print and final error. Scale bar, 2 µm. (a–c,e) Adapted under the terms of the CC BY 4.0 license [5]. Copyright 2026, The Authors, published by Springer Nature. (d) Adapted under the terms of the CC BY 4.0 license [44]. Copyright 2021, The Authors, published by Springer Nature. (f,g) Adapted under the terms of the CC BY 4.0 license [97]. Copyright 2025, The Authors, published by Springer Nature. Original panel labels were replaced.

Measurement can also move into the printer. Zvagelsky et al. added in situ quantitative phase imaging to a commercial printer and recorded a through-focus stack within seconds, instead of optimizing micro-optics ex situ after development [99]. Hu et al. combined video analysis during printing with image analysis afterwards and classified print quality with 93.1% accuracy in 48 ms [100]. Because errors at individual foci must be found and corrected, inspection must scale with the number of foci.

## 11 Applications Enabled by Parallel Writing

Faster writing matters because it expands what TPL can make. Micro-optics, metamaterials, microelectromechanical systems, packaging, biotechnology, quantum technology, and fusion and high-energy laser targets all require both speed and nanoscale precision [5, 46]. This section groups the demonstrations reviewed above by application (Figure 11).

**Mechanical metamaterials.** Architected materials need high voxel counts, because their properties emerge from many unit cells. Fixed arrays printed chiral metamaterials with $3 \times 10^{11}$ and more than $1.7 \times 10^{12}$ voxels (Figure 11a) [22, 23], hybrid arrays printed chainmail, octet and Kelvin lattices [46], and continuous projection produced millimetre-scale lattices (Figures 8d and 11b) [44].

**Photonic structures.** Periodic photonic structures such as woodpiles suit parallel writing well and have been printed with two-plane DOEs [53] and by high-repetition-rate projection [86]. Other examples include terahertz metamaterials [46], spiral arrays with circular dichroism from volumetric holography [67], photonic crystals in yttrium aluminium garnet (Figure 11c) [68] and a $1 \times 4$ $cm^2$ nano-dot array that diffracts white light into structural colour (Figure 11d) [15].

**Micro-optics.** Micro-optical elements are often needed in arrays with sub-wavelength shape accuracy [6]. Parallel methods have produced a $6 \times 5$ lens array in 2 min [38], microlenses in under 120 ms each by continuous projection (Figure 11e) [44], hemispherical microlens arrays by multi-focus holography [59] and uniform micropillar arrays [86]. TPL even prints the DOEs and lens arrays of parallel printers [23], and in situ phase imaging could shorten the print–measure–correct cycle [99].

**Optical computing.** Diffractive processors need millions of precisely placed features. A 1 $mm^2$ element with four million optical neurons, printed in 15 min by randomized multi-focus writing, served as an optical neural network for visible-wavelength machine vision (Figure 11f). In this application, the neuron count, and thus the throughput, directly limits performance [63].

**Cell scaffolds and medical devices.** Cell studies and therapies often require many identical scaffolds. Maibohm et al. printed 81 scaffold units, on which HeLa cells showed elongated morphologies and signs of increased proliferation [8]. Multi-focus Nichoid scaffolds supported stem-cell expansion [58], and four SLM foci produced microneedle arrays and scaffolds [57]. As a serial benchmark, resonant scanning reached 12,078 microscaffolds per hour [40].

**Cell and particle manipulation.** Shaped beams have mainly produced capture devices. Bessel-written microtubes captured and released cells [72], Mathieu-beam microcages trapped micro-objects [73], and autofocusing-vortex microtraps captured microspheres with more than 90% success (Figure 11g) [75]. Single exposures suit such devices because each contains many similar traps.

**Microparticles.** Parallel writing offers a route to the mass production of free-floating microparticles. The 7 × 7 printer made about 7 million carrier particles of three designs for potential pharmaceutical use, 415,373 per 2-inch wafer in 8 h (Figure 11h) [23], and a metalens array made more than 50 million per day [46]. At this scale, TPL begins to move from prototyping towards manufacturing.

**Targets, devices and packaging.** Laser and fusion targets, microelectromechanical systems, packaging and quantum devices [5, 46] remain largely proposed rather than demonstrated, and they will test whether parallel TPL can meet industrial requirements for yield, repeatability and inspection. Smaller actuated devices are closer to use. Ouyang et al. printed magnetic microgears with 32 foci, at about 290 ms per gear, and steered them with a 10 mT field (Figure 11i) [15].

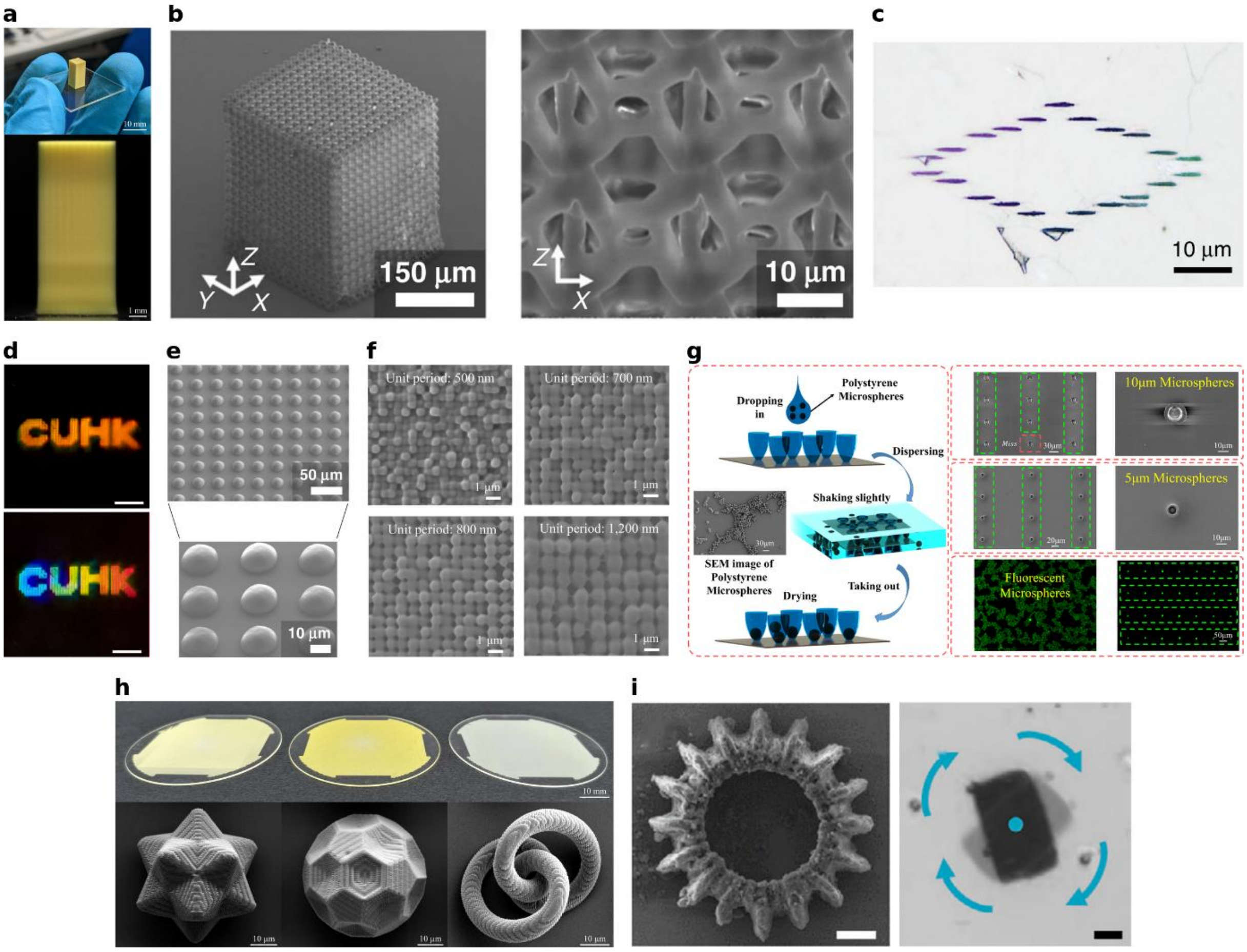

**Figure 11.** Applications of parallel two-photon writing. (a) A 4.62 × 4.62 × 10.26 $mm^3$ chiral metamaterial with more than 1.7 × $10^{12}$ voxels, printed with a 7 × 7 focus array: photograph (top) and optical micrograph (bottom). (b) A 15 × 15 × 15 unit-cell lattice printed by continuous projection, with a magnified side view. (c) Rhombus-like structures written inside a YAG crystal by multi-beam femtosecond processing. (d) A 1 × 4 $cm^2$ "CUHK" pattern of nano-dot arrays with a 4 µm period, written by holographic multi-focus printing, which appears in one or many colours depending on the illumination angle (scale bars, 1 cm). (e) An array of hemispherical microlens-like structures printed by continuous projection, with a magnified view. (f) 3D-printed diffractive layers for an optical neural network, with unit-cell periods of 500–1200 nm. (g) Bowl-shaped microtraps capturing polystyrene microspheres: trapping procedure (left), microspheres captured in traps with 10 and 5 µm openings (top right) and fluorescence images before and after trapping (bottom right). (h) Carrier microparticles for respiratory drug delivery: three 2-inch wafers, each with 415,373 particles (top), and particles of three designs (bottom). (i) A magnetic microgear printed with 32 foci (left, scale bar 10 µm) and a released gear rotating in a magnetic field (right). (a,h) Adapted under the terms of the CC BY 4.0 license [23]. Copyright 2024, The Authors, published by Light: Advanced Manufacturing. (b,e) Adapted under the terms of the CC BY 4.0 license [44]. Copyright 2021, The Authors, published by Springer Nature. (c) Adapted under the terms of the CC BY 4.0 license [68]. Copyright 2023, The Authors, published by Springer Nature. (d,i) Adapted under the terms of the CC BY 4.0 license [15]. Copyright 2023, The Authors, published by Springer Nature. (f) Adapted under the terms of the CC BY 4.0 license [63]. Copyright 2026, The Authors, published by Light: Advanced Manufacturing. (g) Adapted under the terms of the CC BY 4.0 license [75]. Copyright 2023, The Authors, published by MDPI. Original panel labels were replaced.

## 12 Comparison of Approaches

Table 2 summarizes the main families. Its rates are peak values obtained under different voxel definitions, resolutions and test geometries (Section 3), so they indicate scale rather than rank.

**Table 2.** Families of parallel TPL: control of the foci, representative results and main limitations. Rates are peak values under the voxel definitions of the original studies.

| Approach | Control of foci | Representative results | Main limitations |
|---|---|---|---|
| Fixed arrays (microlens arrays, DOEs, metalenses) | All foci identical and scanned together | 3 × 3 DOE, near $10^7$ voxels $s^{-1}$ [22]. 7 × 7 DOE, ≈$10^8$ voxels $s^{-1}$ [23]. Seven-focus metalens, 95.21% uniformity [52] | Periodic or replicated parts only, uniformity fixed by the element and stitching [8] |
| Liquid-crystal SLM holography | Position and intensity of each focus, at tens of hertz | Different objects written at once [56], 16 foci for medical devices [57] and 1.49 × $10^8$ voxels $s^{-1}$ with >400 foci and galvanometric scanning [64] | Slow refresh, low damage threshold [5] and speckle from iterative holograms [66] |
| DMD holography | Binary holograms at kilohertz rates | One to tens of foci at 22.7 kHz [61], and 2000 foci at 2 × $10^6$ voxels $s^{-1}$ with 90 nm resolution [15] | Low light efficiency, high power per focus, speckle [66] and limited packing of foci [5] |
| Hybrid arrays (fixed lenses, programmable illumination) | Each focus switched and graded | Thousands of foci with features below 200 nm [65], and >120,000 foci at >$10^8$ voxels $s^{-1}$ with 113 nm features [46] | Complex optics, and aperiodic parts need pattern compression [46] |
| Shaped beams (Bessel, Mathieu, autofocusing) | One extended focus per exposure, shaped by a hologram | Whole cylinder in 5–10 s [71], microcage in 0.36 s instead of 54 s [73] and 10 × 10 trap array in ≈3 min [75] | Shapes limited to the beam family, side-lobe losses and dose variation along the extended focus |
| Projection with temporal focusing | Whole layer per pulse, with a grayscale motif per spot | ≈3.3 × $10^8$ voxels $s^{-1}$ per projected pixel [19], 1.08 × $10^{-3}$ $mm^3$ $s^{-1}$ in continuous printing [44], 7.9 × $10^7$ or 1.7 × $10^9$ voxels $s^{-1}$ with 55 nm wires [5] and >2.3 × $10^6$ voxels $s^{-1}$ at 500 kHz [86] | High pulse energy, DMD heating [5] and supports for overhangs [15] |
| Two-step absorption with continuous-wave light | DMD holography or two-colour light sheet | 61,500 voxels $s^{-1}$ with 31 foci [42] and 7 × $10^6$ voxels $s^{-1}$ at 0.55 $\mu m^3$ with a light sheet [90] | Coarser light-sheet voxels, and the intermediate-state lifetime caps the single-focus speed [42] |

Table 3 collects representative results for specific systems, as reported.

**Table 3.** Representative parallel TPL systems and their reported performance, in chronological order. Dashes mark values not stated in the cited sources.

| Study | Method | Foci or spots | Light source | Reported rate | Smallest feature or voxel |
|---|---|---|---|---|---|
| L. Yang et al. 2014 [71] | SLM-shaped Bessel beam, single exposure | One shaped focus | — | 5–10 s per cylinder | 1.25–1.56 µm walls |
| C. Wang et al. 2019 [73] | SLM-shaped Mathieu beam | One shaped focus | — | 0.36 s per cage (54 s serial) | — |
| Saha et al. 2019 [19] | Projection with temporal focusing | Up to ≈$10^6$ pixels | 800 nm, 1 kHz, 35 fs | ≈3.3 × $10^8$ voxels $s^{-1}$ (per pixel) and 5–20 $mm^3$ $h^{-1}$ | 130–140 nm wide lines |

| Study | Method | Foci or spots | Light source | Reported rate | Smallest feature or voxel |
|---|---|---|---|---|---|
| Hahn et al. 2020 [22] | 3 × 3 DOE, scanned | 9 | — | Near $10^7$ voxels $s^{-1}$ | Sub-micrometre voxels |
| Somers et al. 2021 [44] | Continuous DMD projection with temporal focusing | One layer | 800 nm, 5 kHz, 65 fs | $3.2 \times 10^6$ voxels $s^{-1}$ and $1.08 \times 10^{-3}$ $mm^3$ $s^{-1}$ | Lines below 200 nm |
| Ouyang et al. 2023 [15] | DMD holography, one pulse per voxel | Up to 2000 | 800 nm, 1 kHz, 100 fs, 4 W | $2 \times 10^6$ voxels $s^{-1}$ | 90 nm |
| Kiefer et al. 2024 [23] | 7 × 7 DOE and lens array, galvanometric scanning | 49 | 790 nm, 80 MHz, 140 fs, 3.7 W | $1.4 \times 10^8$ voxels $s^{-1}$ ($7.1 \times 10^7$ with averaged voxels) | ≈200 nm × 500 nm voxels |
| X. Wang et al. 2024 [52] | Shared-aperture metalens | 7 | — | Seven times single-focus printing | — |
| Zhang et al. 2024 [64] | LCoS-SLM holography with galvanometric scanning | >400 | — | $1.49 \times 10^8$ voxels $s^{-1}$ | — |
| Gu et al. 2025 [46] | Metalens array with SLM illumination | >120,000 | — | $>10^8$ voxels $s^{-1}$ | 113 nm |
| Rietz et al. 2025 [42] | Two-step absorption, DMD holography | Up to 31 | Continuous wave | 61,500 voxels $s^{-1}$ | — |
| K. Liu et al. 2026 [66] | Holograms from model-driven deep learning | Up to 2000 per exposure | — | $1.2 \times 10^5$ voxels $s^{-1}$ | 120 nm |
| L. Xu et al. 2026 [95] | Parallel peripheral photoinhibition | 10 | — | $\approx 10^7$ voxels $s^{-1}$ | 20.2 nm lines |
| Ouyang et al. 2026 [63] | Randomized DMD multi-focus writing | 25 per exposure | 800 nm, 1 kHz, 100 fs | 0.267 million optical neurons per minute | 500 nm pixels |
| Kim and Saha 2026 [5] | Grayscale projection with temporal focusing | 15,836 spots | 804 nm, 5 kHz, 35 fs | $7.9 \times 10^7$ or $1.7 \times 10^9$ voxels $s^{-1}$ and ≈215 $mm^3$ $h^{-1}$ | 55 nm wires |
| Papamakarios et al. 2026, preprint [86] | Projection with spatiotemporal beam shaping | — | 515 nm, 500 kHz, <100 fs | $>2.3 \times 10^6$ voxels $s^{-1}$ | Below 400 nm |

Sorted by light source, Table 3 shows a pattern. A megahertz oscillator fed 49 rapidly scanned foci [23], kilohertz amplifiers fed thousands of holographic foci or a whole projected layer per pulse [5, 15, 19], and continuous-wave sources drove two-step absorption with tens of foci or a light sheet [42, 90]. Because the highest rates, $10^8$–$10^9$ voxels $s^{-1}$, come from distinct architectures, performance now appears to be limited less by the optical architecture than by power, dose control and materials (Sections 9 and 10).

Figure 12 compares the same results along two axes. Most systems published since 2019 exceed $10^6$ voxels $s^{-1}$, and the fastest reach $10^8$–$10^9$ voxels $s^{-1}$ (Figure 12a) [5, 19, 23, 46, 64], with features of about 50–200 nm. The finest parallel features, about 20 nm, came instead from photoinhibition at about $10^7$ voxels $s^{-1}$ (Figure 12b) [95]. Two-step absorption and open-source holography trade throughput or voxel size for lower cost [42, 90, 93], whereas a one-photon holographic exposure, included for comparison, combines high speed with much coarser features [77].

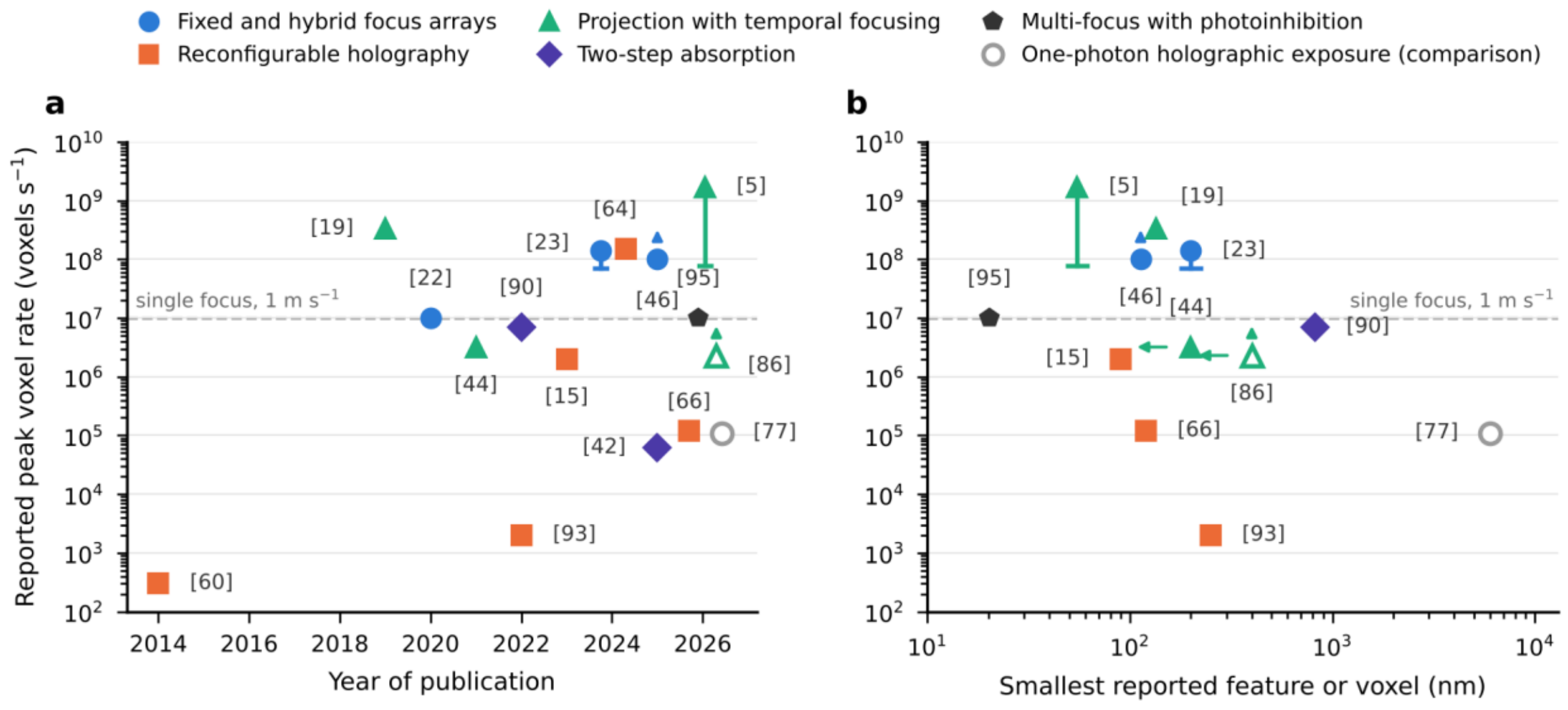

**Figure 12.** Reported throughput of parallel TPL. (a) Peak voxel rate versus year of publication, labelled with reference numbers. (b) Peak voxel rate versus the smallest reported feature or voxel, for the systems that report one. For the light sheet of Hahn et al., the cube root of the 0.55 μm$^3$ voxel is used [90]. Vertical bars extend to a second value reported for the same experiment under another voxel definition [5, 23], and arrows mark lower or upper bounds. The dashed line marks $\approx 10^7$ voxels s$^{-1}$, one focus scanned at 1 m s$^{-1}$ with a 100 nm voxel pitch (Section 2.3). The open triangle marks a preprint [86], and the grey open circle a one-photon holographic exposure shown for comparison [77]. Values are as reported (Table 3 and text), and voxel definitions differ between studies (Section 3), so the plot shows scale rather than a ranking. Original figure prepared for this review from the cited data.

Three patterns stand out. First, no family dominates on every axis. Fixed arrays have produced the largest finished structures, with more than $10^{12}$ voxels [23], but they print only copies. Holography gives the most freedom per focus but pays in efficiency or refresh rate, and projection reports the highest voxel rates but needs amplified lasers and careful dose correction [5]. Second, the families are converging. Hybrid arrays pair fixed lenses with programmable illumination [46, 65], holography borrows fast scanning [64], and projection has adopted per-spot grayscale, the individual control that once defined holography [5]. Third, the right choice depends on the part. Identical microparticles and periodic metamaterials favour fixed and hybrid arrays, and sparse aperiodic parts favour random-access holography [15]. Dense layered structures with fine pores call for grayscale projection [5], many copies of a tube, ring or cage call for a shaped beam [71, 73], and low cost and size favour two-step absorption with continuous-wave diodes [42, 89].

Cost and maturity also separate the families. Fixed arrays run on megahertz oscillators with optics that TPL can itself print [23], whereas the highest-rate holography and projection rely on kilohertz amplifiers [5, 15]. Two-step absorption [89] and open-source holography [93] lower the cost of entry further, and for many laboratories, cost will decide the method before throughput does. Several of the newest results, on multi-plane DOEs, 500 kHz projection and temporal focusing with line illumination, are still preprints [53, 86, 87].

## 13 Outlook and Open Questions

Peak rates of $10^8$–$10^9$ voxels $s^{-1}$ (Section 12) remain far below the $10^{20}$–$10^{21}$ voxels $s^{-1}$ that Somers et al. estimate a single spatiotemporally shaped pulse could, in principle, achieve by printing an entire architecture at once [24]. With the physical ceiling this distant, the central question has shifted from how many voxels can be written simultaneously to how precisely each one can be controlled.

**Shared metrics and benchmarks.** The reported rate of a single experiment can vary about 20-fold with the voxel definition [5]. Beyond the reporting items of Section 3, the field therefore needs shared test parts, such as a lattice with a specified fill factor, a nanoporous block and a solid micro-optic, printed and measured consistently and reported with their energy per voxel and sustained part-level rates.

**Dose prediction for arbitrary geometries.** Per-spot dose control now exists in projection, metalens arrays and holography [5, 15, 46, 66]. The challenge is to predict the right dose for every voxel of an arbitrary part, accounting for proximity effects [7, 25], field non-uniformity and heating [86]. This will likely require physics-based models such as that of Section 2, reaction–diffusion simulations [96], data-driven optimization [97, 98] and in situ metrology [99] working in a closed loop.

**Resolution at speed.** Peripheral photoinhibition narrowed lines to about 20 nm at about $10^7$ voxels $s^{-1}$ with ten channels [95], and co-confinement chemistry reached 30 nm critical dimensions [94]. Whether these approaches can scale to thousands of foci without sacrificing sensitivity remains open.

**Truly 3D parallelism.** Most parallel methods still expose one plane at a time. Volumetric holography [67] and two-plane DOEs [53] point towards simultaneous exposure at multiple depths, with the single-pulse concept as the limiting case [24]. The obstacles are out-of-plane overlap, which grows with the number and density of foci (Section 2.4) [7], and the need for resists with sharp, predictable thresholds.

**Designed light fields.** Shaped beams cure whole tubes, rings or cages [71, 73, 75], but only in established beam families. Inverse design could lift this restriction. In one-photon lithography, inverse-designed masks already project prescribed 3D intensity distributions through thick resist [77, 78]. Extending such computed light fields to two-photon resists, with their sharper thresholds, and replicating them holographically could produce more diverse building blocks in a single exposure.

**Power, cost and resists for parallel writing.** Because each voxel requires a certain energy [21], further gains must come from more power, more sensitive resists [29], cheaper excitation or more efficient use of light [24], as pursued with high-cross-section initiators [15], continuous-wave two-step absorption [42, 88–90] and high-repetition-rate lasers with thermal management [86]. Most resists, however, were developed for single-focus printers. Parallel writing calls for resists that are highly sensitive [15, 29], shrink little [17], tolerate heating [86] and sharpen features without losing sensitivity [94, 95], and, for two-step absorption, have intermediate states short-lived enough not to limit speed [42]. No resist combines all these properties, so materials development is as important as optical design. Measuring the nonlinear absorption coefficients and thresholds of candidate resists under the focusing and pulse conditions of the target printer, as in Section 2, offers a practical screening route [35].

**Learning-based design and control.** Machine learning now computes holograms with less speckle [66], tunes projection printing within a few iterations [97, 98] and classifies print quality in tens of milliseconds [100]. Combined with in situ imaging [99] and physical models of proximity effects [46, 96], it could let printers learn from each print and correct the next dose map, provided that models transfer between resists, objectives and geometries and that thousands of printed features can be measured rapidly.

**Data pipelines.** A printer writing $10^9$ voxels $s^{-1}$ must receive data at a comparable rate. Pattern streaming [87], hologram computation [60, 66] and pattern compression [46] are becoming integral to printer design. Converting CAD models into dose-corrected patterns for thousands of foci fast enough to keep the laser busy, however, remains an open problem.

**Hybrid architectures.** Leading systems already combine approaches, such as holography with galvanometric scanning [64], metalens arrays with SLM illumination [46] and DMD line illumination with continuous stage motion [87]. Future printers may pair a fixed element that sets the number and spacing of the foci with a modulator that sets their doses and a scanner that sets the speed, optimized as one system together with the laser and resist.

**From demonstrations to production.** Metalens arrays target wafer-scale manufacturing [46], projection has reached 500 kHz [86], and resonant scanning has raised even serial writing to thousands of scaffolds per hour [40]. Yield, repeatability and cost per part over extended production runs, which are still rarely reported, will decide which approaches move from the laboratory to manufacturing.

## 14 Conclusions

TPL combines true 3D design freedom with features far below the wavelength of light, but voxel-by-voxel writing kept it slow for more than two decades. Parallel writing has changed this, with more than $10^5$ simultaneous foci and peak rates of $10^8$–$10^9$ voxels $s^{-1}$, depending on the architecture and the voxel definition. Four conclusions emerge.

First, the throughput ceiling is set by the energy per voxel, not by optics alone. Every focus must exceed the curing threshold, so laser power, resist sensitivity and optical efficiency are central to scalability. The framework of Section 2 quantifies these dependencies and the added costs of parallel writing in uniformity, crosstalk, side lobes and axial confinement. These predictions rest on the absorption order and coefficients of the resist, which, combined with one measured threshold, show which writing regimes and parallel architectures a given resist can support (Section 2.5).

Second, each approach serves a distinct purpose: fixed and hybrid arrays for replicated parts, holography for sparse aperiodic structures, shaped beams for tubes, rings and cages, projection for dense layers with per-spot grayscale, and two-step absorption for low-cost printing with continuous-wave lasers.

Third, the bottleneck has shifted from speed to control. Non-uniformity, proximity effects, heating, stitching and data handling now limit quality and scale, and the best responses so far are grayscale compensation, model-based and data-driven dose correction, and in situ measurement.

Fourth, the field needs common benchmarks: explicit voxel definitions, resolution metrics and shared test parts, together with the energy per voxel and sustained part-level rates that headline rates do not capture.

From metamaterials, micro-optics and optical computing to cell scaffolds, microparticles and laser targets, applications require both speed and precision. Further progress will therefore depend less on adding foci than on controlling each one accurately, economically and at scale.

## Acknowledgements

During the preparation of this review, the authors used Claude (Anthropic) to assist with summarizing source literature and improving language clarity. All summaries, data, and interpretations were independently verified against the original publications by the authors, who takes full responsibility for the final content.

## Conflicts of Interest

The authors declare no conflicts of interest.

## Data Availability Statement

This review reports no new experimental data. All values in Tables 1–3 and Figure 12 are taken from, or calculated from, the cited publications. The scripts used for the calculations in Section 2 and to produce Figures 3 and 12 are available from the corresponding author upon reasonable request.